\documentclass[twocolumn,aps,prb,showpacs,amsmath,superscriptaddress]{revtex4}
        \usepackage{amsmath}
        \usepackage{amssymb}
        \usepackage{natbib}
        \usepackage{epsfig}
        \usepackage{amsfonts}
        \usepackage{mathrsfs}
        \usepackage{color}
        \usepackage{bm}
        \usepackage[toc,page,title,titletoc,header]{appendix}
        \usepackage{CJK}
        \usepackage{epstopdf}
    \usepackage{mathrsfs}
    \usepackage{float}
    \usepackage{tikz-cd}
    \usepackage{graphicx}
    \usepackage{dcolumn}
    \usepackage{bm}
\usepackage{multirow}
\usepackage{comment}
\usepackage{indentfirst}
\usepackage{slashed}

\begin{document}
      \widetext
     \title{Information constraint in open quantum systems}
      \author{Chun-Hui Liu}
       \affiliation{Beijing National Laboratory for Condensed Matter Physics, Institute of Physics, Chinese Academy of Sciences, Beijing 100190, China}
\affiliation{School of Physical Sciences, University of Chinese Academy of Sciences, Beijing 100049, China}
      \author{Shu Chen}
      \email{schen@iphy.ac.cn}
\affiliation{Beijing National Laboratory for Condensed Matter Physics, Institute of Physics, Chinese Academy of Sciences, Beijing 100190, China}
\affiliation{School of Physical Sciences, University of Chinese Academy of Sciences, Beijing 100049, China}
\affiliation{Yangtze River Delta Physics Research Center, Liyang, Jiangsu 213300, China}
          \begin{abstract}
           \par
           We propose an effect called information constraint which is characterized by the existence of different decay rates of signal strengths propagating along opposite directions. It is an intrinsic property of a type of open quantum systems, which does not rely on boundary conditions. We define the value of information constraint ($I_C$) as the ratio of different decay rates and  derive the analytical representation of $I_C$ for general quadratic Lindbladian systems. Based on information constraint, we can provide a simple and elegant explanation of chiral and  helical damping,  and get the local maximum points of relative particle number for the periodical boundary system, consistent with numerical calculations. Inspired by information constraint, we propose and prove the correspondence between edge modes and damping modes. A new damping mode called Dirac damping is constructed, and chiral/helical damping can be regarded as a special case of Dirac damping.

          \end{abstract}
          \maketitle

          \section{Introduction}
          Many open quantum systems can be effectively described by non-Hermitian matrix or Hamiltonian, e.g., the short time evolution of the Lindblad master equation Eq.(\ref{lindbladeq}) is governed by non-Hermitian Hamiltonian $H_{NH}=H-i\sum_{\mu}L_{\mu}^{\dagger}L_{\mu}$  before the occurrence of first quantum jump \cite{GLindblad, Molmer, Carmichael, Daley}. Essentially, the Lindblad master equation can be mapped to a non-Hermitian ``Sch\"{o}rdinger equation" even with quantum jump term after using a basis to represent the density matrix \cite{TProsen1, TProsen2, TProsen3,SLieu,Zoller}, and the calculating the Lindbladian spectrum of the superoperator can always be viewed as a non-Hermitian eigenvalue problem. Particles with finite lifetime can also be effectively described by non-Hermitian Hamiltonian\cite{HShen,VKozii}. Non-Hermitian systems have been unveiled to possess  some unique features, such as non-Hermitian skin effect (NHSE) \cite{Alvarez,SYao1,SYao2,TELee,Kunst,KZhang,Xiong,KYokomizo,LeeCH,JiangH,Slager,NOkuma,ZSYang,LHLi,LeeCH2019,YYi}, exceptional points, \cite{Heiss,Dembowski,Rotter,Hu2017,Hassan2017,Kim,LeiPan} and amplified symmetry classes \cite{Gong,Sato,Zhou,CHLiu1,CHLiu2,Ueda}. These unique features produce significant influence on the time evolution of the system and give rise to some peculiar dynamical phenomena, such as chiral/helical damping for non-Hermitian skin effect \cite{FSong1,CHLiu3} and amplifying sensors for exceptional points \cite{Hodaei,ChenW,Wiersig1,Wiersig2}.

          The NHSE relies on the boundary condition, and no NHSE and non-Bloch wave can be observed for systems under the periodic boundary condition (PBC), while Bloch's theorem is valid under the PBC.
          Similarly, the phenomena of chiral and helical damping in open quantum systems occur only under the open boundary condition (OBC). An important issue is to extract the intrinsic property\cite{IP} for systems exhibiting NHSE and get a unique feature which is not sensitive to boundary conditions. We expect that this feature can explain chiral/helical damping without resorting to NHSE.

          In this work, we propose an effect called information constraint
          characterized by the existence of different decay rates of signal strengths  propagating along opposite directions, which induces the information propagation being constrained in one of directions.
          The ratio of strengths propagating along opposite directions, or equivalently the ratio of local two-point Green functions along opposite directions, defines the value of information constraint $I_C$.  Since $I_C$ is a local quantity, its value should not rely on boundary condition, which allows us to derive $I_C$ by using arbitrary boundary condition. Under the PBC, we are able to analytically calculate local maximum points of relative particle number via information constraint, which show obviously different distributions along different propagation directions and are consistent with numerical results. Based on information constraint, we get a simple and elegant explanation of chiral and helical damping, and deduct naturally  the helical damping model supporting the helical tunneling effect \cite{CHLiu3,YYi}.
          Inspired by information constraint, we propose and prove the correspondence between edge modes and damping modes. A new damping mode called Dirac damping is constructed as an example, with chiral/helical damping as a special case of one-dimensional (1D) Dirac damping.
          \begin{figure*}[t]
\includegraphics[width=1\linewidth]{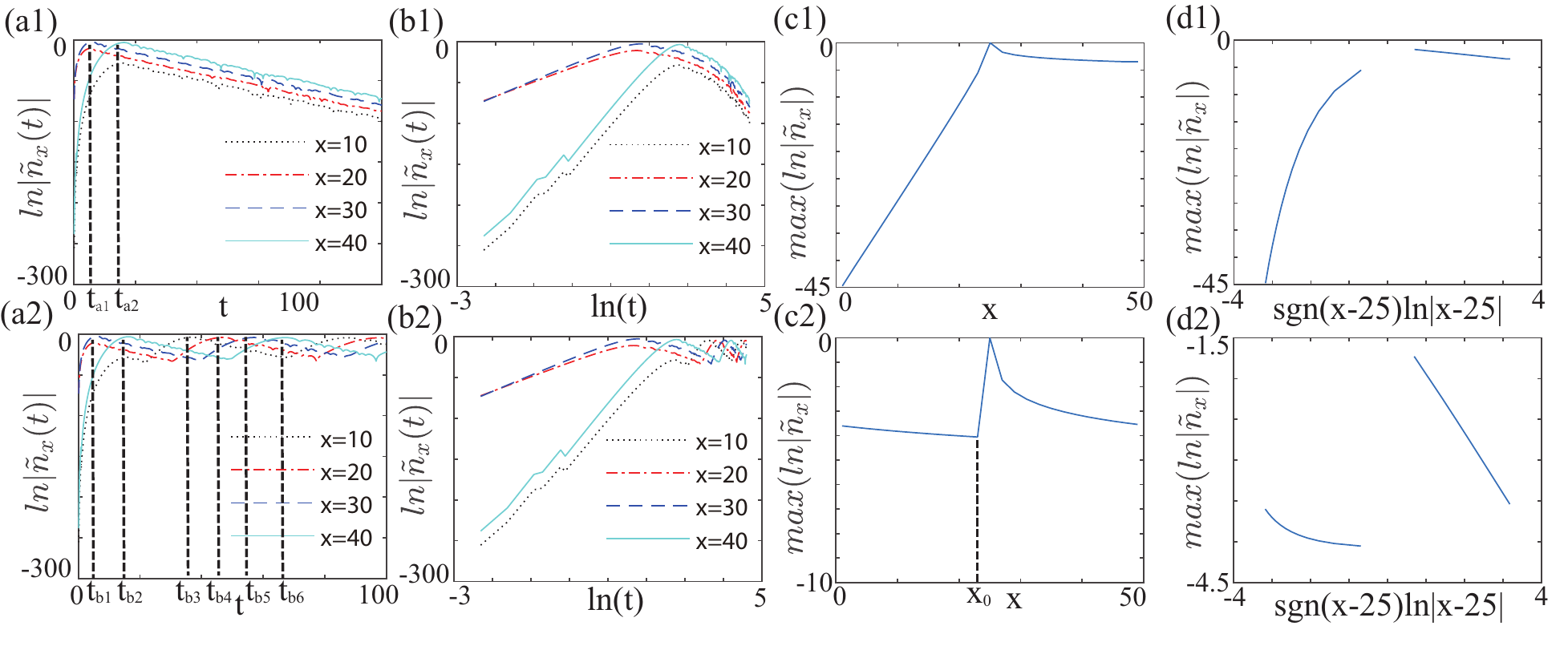}\\
\caption{(a1)--(d1) are for the system under OBC, and (a2)--(d2) for PBC. The parameters are $t_1=t_2=1$ and $\gamma=0.8$. [(a1) and (a2)] The evolution of $ln|\tilde n_x(t)|$ as a function of $t$ at cells $x=10$, $20$, $30$, and $40$. In (a1), $t_{max}(20)=t_{max}(30)=t_{a1}$ and $t_{max}(10)=t_{max}(40)=t_{a2}$. In (a2), $t_{b1}=5$, $t_{b2}=15$, $t_{b3}=35$, $t_{b4}=45$, $t_{b5}=55$ and $t_{b6}=65$. Here $t_{b2}$ and $t_{b3}$ are local maximum points of $ln|\tilde n_{10}(t)|$, $t_{b1}$ and $t_{b4}$ are of $ln|\tilde n_{20}(t)|$, $t_{b1}$ and $t_{b5}$ are of $ln|\tilde n_{30}(t)|$, and $t_{b2}$ and $t_{b6}$ are of $ln|\tilde n_{40}(t)|$. [(b1) and (b2)] The evolution of $ln|\tilde n_x(t)|$ as a function of $ln(t)$.  The straight line indicates that $|\tilde n_x(t)|$ is a power-law function of $t$ at corresponding interval $[0,t_{max}(x)]$. [(c1) and (c2)] $max(ln|\tilde n_x|)$ as a function of $x$.  In (c2), if $x\in[x_0,25]$, $max(ln|\tilde n_x|)$ is dominated by signal from the $-x$ direction (the exponential decay mode). If $x\in[1,x_0]\cup[25,50]$, $max(ln|\tilde n_x|)$ is dominated by signal from the $+x$ direction (the power-law decay mode). Thus, it is not an analytical function at $x_0$. [(d1) and (d2)] $max(ln|\tilde n_x|)$ as a function of ${\bm sgn(x-25)ln|x-25|}$. The straight line indicates that $max(|\tilde n_x|)$ is a power-law function of $x-25$ at $x\in[25,50]$.} \label{fig1}
\end{figure*}

     \section{Information constraint}
     To illustrate the concept of information constraint, we study the particle transport in open 1D chains and demonstrate that the information constraint is an intrinsic property of a type of open quantum systems. Consider the open Markovian quantum systems described by the Lindblad master equation
      \begin{equation}
    \frac{d\rho}{dt} 
    =\mathcal{L}[\rho]=-i[H,\rho]+\sum _{\mu}\left(2L_{\mu}\rho L_{\mu}^{\dagger}-\left\{L_{\mu}^{\dagger}L_{\mu},\rho\right\}\right) ,
    \label{lindbladeq}
    \end{equation}
where $\rho$ is the density matrix, $L_{\mu}$ are Lindblad operators describing quantum jump processes, and $H$ is the Hamiltonian. To make concrete, we consider the Su-Schrieffer-Heeger (SSH) model with the Hamiltonian in the momentum space given by
\begin{equation}
   \begin{split}
    h(k)=[t_1+t_2 \cos (k)]\sigma_x+t_2 \sin (k)\sigma_y ,
    \end{split}
    \label{operators2}
    \end{equation}
where $t_1$ and $t_2$ represent the hopping amplitude in and between the unit cells, respectively, and there are two ($A$ and $B$) sublattices in each cell. While $\sigma_0$ denotes a $2\times2$ identity matrix, $\sigma_x$, $\sigma_y$ and $\sigma_z$ represent Pauli matrices. The coupling to the environment is described by the Lindblad operators given by
  \begin{equation}
   \begin{split}
    L_{x}^l=\sqrt{\frac{\gamma}{2}}(c_{xA}-ic_{xB}) ,
    \end{split}
    \label{operators1}
    \end{equation}
  where $c_{xA}$ and $c_{xB}$ are fermion annihilation operators on the site $xA$ and $xB$, respectively, and  $x$ is the cell index \cite{xindex}. The dynamics of
  \[
  \Delta_{mn}=Tr(\rho c_m^{\dagger}c_n)
  \]
  with $m,n\in \left\{xA,xB\right\}$ is governed by \cite{Zoller,TProsen1,FSong1,CHLiu3}
    \begin{equation}
   \tilde{\Delta}=\Delta-\Delta_s=e^{Xt}\tilde{\Delta}(0)e^{X^{\dagger}t},
   \end{equation}
   where $\Delta_s$ is the steady value of $\Delta$ ($\Delta_s=0$ for this model) and $X$ is a damping matrix with the matrix in the momentum space given by
    \begin{equation}
   X(k)=ih^{T}(-k)+\frac{\gamma}{2}\sigma_y-\frac{\gamma}{2}\sigma_0.
   \label{Xk}
   \end{equation}
   The diagonal elements of $\tilde{\Delta}$ give the relative particle number defined by $\tilde n_x(t)=\tilde{\Delta}_{xA,xA}+\tilde{\Delta}_{xB,xB}$.

  Suppose that  a particle is initially prepared at the site $25B$ and the system size is 50, and we have $\tilde{\Delta}_{25B,25B}(0)=1$ and $\tilde{\Delta}_{m,n}(0)=0$ when $m$ or $n$ $\ne$ $25B$.  It can be recognized as a signal initially input at site $25B$. We numerically calculate $ln|\tilde n_x(t)|$ versus $t$ or $ln(t)$ under both OBC and PBC in Figs.\ref{fig1}(a1),\ref{fig1}(b1) and \ref{fig1}(a2),\ref{fig1}(b2). Figures.\ref{fig1}(a1) and \ref{fig1}(b1) show the evolution of $ln|\tilde n_x(t)|$ at $x=10$,  $20$, $30$ and $40$ under OBC. For a fixed $x$, $|\tilde n_x(t)|$ increases in a power-law to the maximum value $max(|\tilde n_x|)$ at $t_{max}(x)$ (In the main text, $max()$ is the label of $max_t()$, which is the maximum over all possible time interval), and exponentially decreases after $t_{max}(x)$. The $t_{max}(x)$ can be recognized as the time when the signal reaches $x$ (the location of wave front), and $max(|\tilde n_x|)=exp[max(ln|\tilde n_x|)]$ is the signal strength for the case of OBC \cite{nmax}. As $max(ln|\tilde n_x|)$ is a single-value function of $x$, we illustrate it in Figs.\ref{fig1}(c1) and \ref{fig1}(d1) for OBC and  Figs.\ref{fig1}(c2) and \ref{fig1}(d2) for PBC. While the signal strength decreases exponentially when propagating along the $-x$ direction ($x\rightarrow x-1$), it exhibits  a power-law decay  along the $+x$ direction ($x\rightarrow x+1$). The signal strength has different decay rate in the opposite direction, and we dub this phenomenon as {\it information constraint},  since the information propagation is constrained in one of directions. A quantitative definition of information constraint by using local Green function will be given by Eqs.(\ref{ir}) and (\ref{ipm2}).

   The decay rate and local Green function are both local function and only rely on local dynamical property \cite{PeskinAndSchroeder}. If the Lindblad equation Eq.(\ref{lindbladeq}) is a local equation, i.e., without any long-range coupling in Eq.(\ref{lindbladeq}), the local dynamical property should not rely on boundary condition. Thus, we have the following proposition.

   {\bf Proposition I:} Information constraint does not rely on boundary condition.

      With the increase in time, the evolution of $ln|\tilde n_x(t)|$ (or $|\tilde n_x(t)|$) has many local maximum points under the PBC in Fig.\ref{fig1}(a2). For $x=10$ or $20$, the local maximum points are found at $t_{locmax}=25-x$ and $x+25+50N$, respectively, where $N\ge0$ is an integer. For $x=30$ and $40$, the local maximum point is at $t_{locmax}=x-25+50N$.
    This can be understood in terms of information constraint: If $x \in [1,25]$,  the signal propagating along the $-x$ direction reaches $x$ at time $t_-=\frac{25-x+50N}{v}$, where $v$ is the maximum Fermi velocity of $iX(k)$. Meanwhile, the signal along the $+x$ direction reaches $x$ at time $t_+=\frac{x+25+50N}{v}$. The signal strength at $x$ is dominated  by  the signal from the $+x$ direction after $t_0=t_+|_{N=0} {\bm=} \frac{x+25}{v}$, because the strength of signal from the $+x$ direction exhibits a power-law decay whereas from the $-x$ direction an exponential decay. We analytically calculate the maximum Fermi velocity of $iX(k)$ and get $v=1$. The local maximum points of $ln|\tilde n_x(t)|$ in Fig.\ref{fig1}(a2) come from the signals arriving in $x$. Taking account of $v=1$ and the fact that the signal from the $+x$ direction is dominated after $t_0$, we get the local maximum points at $t_{locmax}=25-x$ and $x+25+50N$ for $x=10$ or $20$.

If $x \in [25,50]$,  the signal propagating along the $-x$ direction reaches $x$ at time $t_-=\frac{75-x+50N}{v}$, whereas the signal along the $+x$ direction reaches $x$ at time $t_+=\frac{x-25+50N}{v}$. The signal strength at $x$ is dominated  by the signal from the $+x$ direction after $t_0=t_+|_{N=0}{\bm =} \frac{x-25}{v}$.
Due to $v=1$, we get the local maximum point at $t_{locmax}=x-25+50N$. The results are consistent with Fig.\ref{fig1}(a2).

  The information constraint can provide a simple and elegant way to understand the chiral and helical damping. For a system under the OBC with size $L$ and ${\forall}x'\in [1,L]$, supposing that the system  is fully filled at the initial time and $v=1$, the particle propagating along the $-x$ direction decays exponentially, whereas the particle propagating along the $+x$ direction decays in power-law. The particles which always propagating along the $+x$ direction will arrive in the cell $x'$ at time $t\in (0,x')$, and these particles contribute a power-law decay factor of $\tilde n_x$. Thus, for $t\in (0,x')$, $\tilde n_x$ decays in a power law.  After $t=x'$, no particle always propagating from the $+x$ direction will arrive at $x'$, and the decay behavior follows a usual relaxation law:
  \[
  \tilde{n}_{x'}(t)\propto e^{-|\Lambda_g|t},
  \]
  where $\Lambda_g$ is the largest non-zero eigenvalue of open boundary Liouvillian superoperators (Liouvillian gap). The combination of $t\in (0,x')$ and $t>x'$ gives rise to chiral damping \cite{FSong1}. In the helical damping case, there are two channels labeled as $\alpha_1$ and $\alpha_2$. In the $\alpha_1$ channel, particle propagating along the $-x$ direction is exponentially decaying, and particle propagating along $+x$ direction exhibits a power-law decay behavior. On the other hand, the decay behavior of the $\alpha_2$ channel  is opposite to that of $\alpha_1$ channel since it fulfills time-reversal symmetry \cite{CHLiu3}. Thus, chiral damping occurs in $\alpha_1$ and $\alpha_2$ channels with wave fronts having opposite propagation directions. The combination of $\alpha_1$ and $\alpha_2$ channels gives rise to helical damping \cite{CHLiu3}.  

     In the $\alpha_1$ channel, the decay rate along  the $+x$ direction is smaller than the $-x$ direction, and thus tunneling amplitude along the $+x$ direction is bigger than the $-x$ direction. This induces chiral tunneling for the $\alpha_1$ channel. Similarly, in the $\alpha_2$ channel, the tunneling amplitude along the $-x$ direction is bigger than the $+x$ direction. Since the two channels have opposite spins (because of time-reversal symmetry \cite{CHLiu3,YYi}), we get that helical tunneling must exist in the helical damping model \cite{helicaltunneling, CHLiu3}. In the Appendix A, we derive tunneling amplitude for the helical damping model, and show the helical tunneling behavior.
     \begin{figure}[h]
\includegraphics[width=0.8\linewidth]{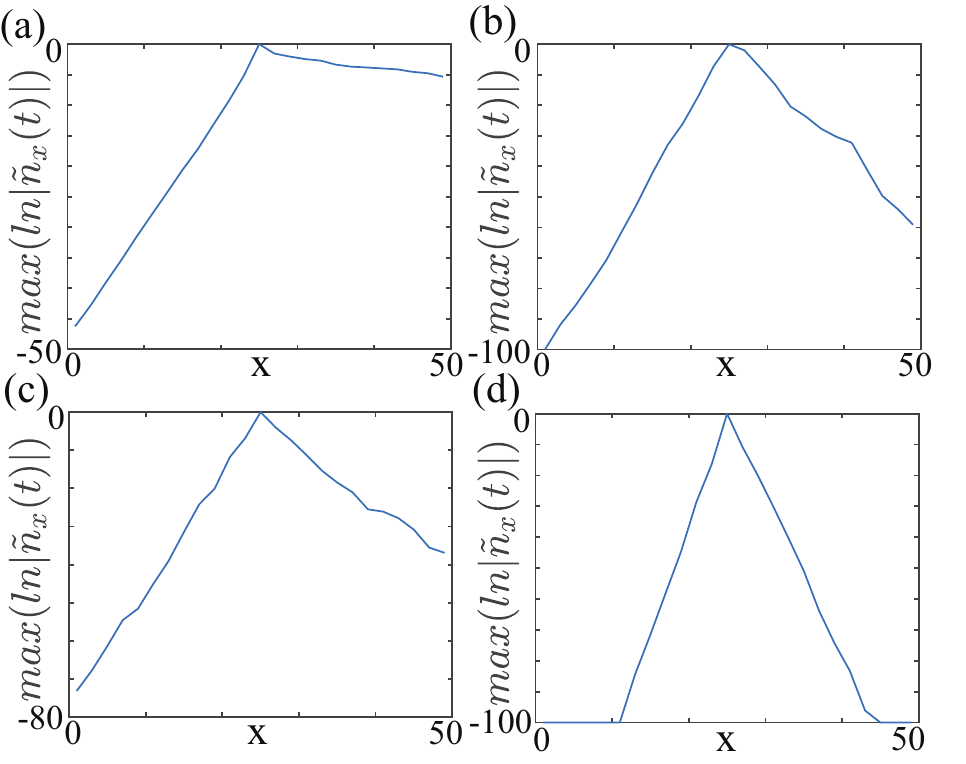}\\
\caption{$max(|\tilde n_x|)$ as a function of $x$ under OBC. (a) and (b) correspond to the model of  Eqs.(\ref{operators1}) and (\ref{perturbation1}) with disorder strength  $W=1$ and $W=10$, respectively. (c) and (d) correspond to the model of Eqs.(\ref{operators1}) and (\ref{perturbation2}) with disorder strength $W=10$ and $W=100$, respectively. The other parameters are $t_1=t_2=1$ and $\gamma=1.0$.} \label{fig2}
\end{figure}

Now we study the effect of disorder and illustrate that the information constraint is stable against disorder. We first consider the disorder introduced in the hopping amplitude with the Hamiltonian in the Lindblad  equation described by
 \begin{equation}
    H=\sum_{i=1}^N[t_1c_{i,A}^{\dagger}c_{i,B}+(t_2+Wr_i)c_{i,B}^{\dagger}c_{{i+1},A} +h.c.],
    \label{perturbation1}
 \end{equation}
 and the Lindblad operators given by Eq.(\ref{operators1}), where $r_i\in (0,1)$ is a random variable and $W$ is the strength of disorder.
 The initial state is taken the same as the case in the absence of disorder.
  We numerically calculate $max(\tilde n_x)$ under the OBC, which is illustrated in Fig.\ref{fig2}(a) and Fig.\ref{fig2}(b).  It is shown that the information constraint exists  for both $W=1$ and $W=10$.  Then we consider random disorder in the chemical potential with the Hamiltonian described by
   \begin{equation}
   \begin{split}
    H=\sum_{i=1}^N[t_1c_{i,A}^{\dagger}c_{i,B} +t_2 c_{i,B}^{\dagger}c_{{i+1},A}+h.c.+Wr_ic_{i,A}^{\dagger}c_{i,A}]
    \end{split}
    \label{perturbation2}
 \end{equation}
and the Lindblad operators are the same as Eq.(\ref{operators1}). Similarly, we numerically illustrate  $max(\tilde n_x)$ under the OBC  in Fig.\ref{fig2}(c) and \ref{fig2}(d). The information constraint exists for $W=1$. With the increase of $W$, the information constraint is suppressed, but there still exists signature of different decaying rates in different propagating directions even for $W=100$. Our results indicate that the information constraint is robust against the disorder.

We note that information constraint also exists in the open spin systems. An example of 1D open Heisenberg XX spin chain is given in the Appendix B, where we show the existence of information constraint by transforming the spin model to a quadratic fermion model.

\section{Correspondence between edge modes and damping modes}
 The information constraint not only exists in the quadratic Lindbladian system and leads to chiral and helical damping, it also exists in the anomalous edge modes of topological insulators, e.g., the chiral edge modes of integer quantum Hall effect \cite{Haldane}. It is natural to ask whether there exists a relation between edge modes in topological insulators/superconduators \cite{Haldane,Kane,Bernevig,Fu} and damping modes in the quadratic Lindbladian system \cite{FSong1,CHLiu3}? Here, we give a correspondence between them.

{\bf Proposition II:} For a $d$-dimensional anomalous boundary state of a $(d+1)$-dimensional Hermitian system (topological insulators/superconductors) in symmetry class $s$, there exists a $d$-dimensional quadratic Lindbladian system \cite{QLS} with damping matrix multiplying $i$ belonging to class $s^{\dagger}$, and its damping wave front has the same structure as the dispersion relation of the anomalous boundary state. Here, the damping wave fronts is defined as  the boundary of two regions with different decay or gain rates, and is a $d$-dimensional surface in $(d+1)$-dimensional space-time $({\bf x},t)$. The dispersion relation is a $d$-dimensional surface in $(d+1)$-dimensional momentum-energy $({\bf k},E)$, and $s$ ($s^{\dagger}$) is a label of 10-fold AZ (AZ$^\dagger$) class (See the Appendix C for the introduction of Hermitian and non-Hermitian symmetry class).

Next we give proof of this proposition.
 Consider a $d$-dimensional anomalous boundary state of a $(d+1)$-dimensional Hermitian system in symmetry class $s$. Suppose that the boundary state is  characterized by the following Dirac Hamiltonian,
    \begin{equation}
   \begin{split}
   H_{D}({\bf k})=k_1\Gamma_1+k_2\Gamma_2+...+k_d\Gamma_d,
         \end{split}  \label{HDirac}
    \end{equation}
    where $\left\{\Gamma_i,\Gamma_j\right\}=\delta_{ij}$ and ${\bf k}=(k_1,k_2,...,k_d)$. We can construct the damping matrix of the corresponding quadratic Lindbladian system (damping matrix multiply $i$ belongs to the symmetry class $s^{\dagger}$) as,
      \begin{equation}
   \begin{split}
   iX({\bf k})=&[\sin(k_1)\Gamma_1+ \sin(k_2)\Gamma_2+...+ \sin(k_d)\Gamma_d ] \\
    &+i [ \cos(k_1)+ \cos(k_2)+...+ \cos(k_d)+E_B ]\mathbb{I},
         \end{split}  \label{DM}
    \end{equation}
   where $E_B\le-d$ is a constant, $\mathbb{I}$ is an identity matrix and $X({\bf k})$ is the damping matrix.
     Assume that the dimension of $\Gamma_i$ and $\mathbb{I}$ is $n_{0i}$.  Let $E_B=-d$, the corresponding quadratic Lindbladian system is described by the Hamiltonian
        \begin{equation}
    h({\bf k})= \sin(k_1)\Gamma_1+ \sin(k_2)\Gamma_2+...+ \sin(k_d)\Gamma_d
    \end{equation}
    and the  Lindblad operators
    \begin{equation}
    \begin{split}
    & L_{(x_1,x_2,...,x_d)1p_1p_2...p_{n_{0i}}} \\
    &=2^{-n_{0i}/2}\left[(-1)^{p_1}c_{(x_1,x_2,...,x_d)1}+(-1)^{p_2}c_{(x_1,x_2,...,x_d)2}+... \right.\\
    &+(-1)^{p_{n_{0i}}}c_{(x_1,x_2,...,x_d)n_{0i}}-(-1)^{p_1}c_{(x_1+1,x_2,...,x_d)1} \\
    &\left. -(-1)^{p_2}c_{(x_1+1,x_2,...,x_d)2}-...-(-1)^{p_{n_{0i}}}c_{(x_1+1,x_2,...,x_d)n_{0i}}\right],\\
    & L_{(x_1,x_2,...,x_d)2p_1p_2...p_{n_{0i}}}\\
    &=2^{-n_{0i}/2}\left[(-1)^{p_1}c_{(x_1,x_2,...,x_d)1}+(-1)^{p_2}c_{(x_1,x_2,...,x_d)2}+...\right.\\
    &+(-1)^{p_{n_{0i}}}c_{(x_1,x_2,...,x_d)n_{0i}}-(-1)^{p_1}c_{(x_1,x_2+1,...,x_d)1} \\
    &\left.-(-1)^{p_2}c_{(x_1,x_2+1,...,x_d)2}-...-(-1)^{p_{n_{0i}}}c_{(x_1,x_2+1,...,x_d)n_{0i}}\right],\\
    &~~~~~~\vdots\\
    & L_{(x_1,x_2,...,x_d)dp_1p_2...p_{n_{0i}}}\\
    &=2^{-n_{0i}/2}\left[(-1)^{p_1}c_{(x_1,x_2,...,x_d)1}+(-1)^{p_2}c_{(x_1,x_2,...,x_d)2}+...\right.\\
    &+(-1)^{p_{n_{0i}}}c_{(x_1,x_2,...,x_d)n_{0i}}-(-1)^{p_1}c_{(x_1,x_2,...,x_d+1)1} \\
    &\left.-(-1)^{p_2}c_{(x_1,x_2,...,x_d+1)2}-...-(-1)^{p_{n_{0i}}}c_{(x_1,x_2,...,x_d+1)n_{0i}}\right],
    \end{split}
    \end{equation}
    where $c_{(x_1,x_2,...,x_d+1)j}$ $(j=1,2,...,n_{0i})$ are annihilation operators, ${\bf x}=(x_1,x_2,...,x_d)$ is cell index, and $1,2,...,n_{0i}$ in $c_{(x_1,x_2,...,x_d+1)1}$, $c_{(x_1,x_2,...,x_d+1)2}$, $...$, $c_{(x_1,x_2,...,x_d+1)n_{0i}}$ are the indexes labeling the degree of freedom in the cell.  In the Lindblad operators $L_{(x_1,x_2,...,x_d)m p_1p_2...p_{n_{0i}}}$, $ m=1,2,...,d,\quad p_1=0$ and $p_2,p_3,...,p_{n_{0i}}=0$ or $1$. Thus, there are total $d\times2^{n_{0i}-1}$ Lindblad operators for fixed $(x_1,x_2,...,x_d)$.

    Now we prove that this model satisfies proposition II.  Assume that $E_{\alpha}({\bf k})$ is the eigenvalue of $iX({\bf k})$, $\alpha$ is the band index and $E_{\alpha}({\bf k})$ satisfies $Im(E_{\alpha}({\bf k}))\le 0$. The dynamic of this model is dominated by the longest life-time (maximum imaginary eigenvalue) mode. At ${\bf k}_0=(0,0,..,0)$, $Im(E_{\alpha}({\bf k}))$ takes the maximum value. Expanding Eq.(\ref{DM}) at ${\bf k}_0$ \cite{JYLee}, we get
     \begin{equation}
   \begin{split}
   iX_{eff}({\bf k})=k_1\Gamma_1+k_2\Gamma_2+...+k_d\Gamma_d+i(d+E_B)\mathbb{I}.
         \end{split} \label{DMeff}
    \end{equation}
    The effective theory is the same as Eq.(\ref{HDirac}). And the damping wave front should have the same behavior as Eq.(\ref{HDirac}).

   To display this more explicitly, we consider this model with infinite system size (infinite system size means that the system size is large enough that we do not need to consider the boundary effect at the considered time scale), and it is fully filled in $d$-dimensional disk $D^d$ ($x_1^2+x_2^2+...+x_d^2<R^2$, where ${\bf x}=(x_1,x_2,...,x_d)$ is the coordinate and $R$ is the radius) and empty in $x_1^2+x_2^2+...+x_d^2>R^2$.  $E_{\alpha}^{eff}$ is the eigenvalue of $iX_{eff}({\bf k})$. We only consider the damping behavior in the $D^d$, and there are two possible cases to be considered:

     (1) ${\bf iX_{eff}( {\bf k})\ne k_1\mathbb{I}+i(d+E_B)\mathbb{I}}$. \\
     If $iX_{eff}( {\bf k})\ne k_1\mathbb{I}+i(d+E_B)\mathbb{I}$, we have
     \begin{widetext}
     \begin{eqnarray*}
     v_{eff}&=&\left(\frac{\partial(Re(E_{\alpha}^{eff}({\bf k})))}{\partial k_1},\frac{\partial(Re(E_{\alpha}^{eff}({\bf k})))}{\partial k_2},
    ...,\frac{\partial(Re(E_{\alpha}^{eff}({\bf k})))}{\partial k_d} \right)\\
    &=& \pm \left( \frac{k_1}{\sqrt{k_1^2+k_2^2+...+k_d^2}},\frac{k_2}{\sqrt{k_1^2+k_2^2+...+k_d^2}},...,\frac{k_d}{\sqrt{k_1^2+k_2^2+...+k_d^2}} \right).
    \end{eqnarray*}
    \end{widetext}
    Substituting it and Eq.(\ref{DMeff}) into the equation as follows
      \begin{equation}
   \begin{split}
  \tilde{\Delta}(t)\approx e^{X_{eff}t}\tilde{\Delta}(0)e^{X_{eff}^{\dagger}t},
         \end{split} \label{GF}
    \end{equation}
      we get that the wave front after time $t_0$ is a sphere $S^{d-1}$ with radius $|R-t_0|$ and center at ${\bf x}_0=(0,0,...,0)$. The wave front has a Dirac cone structure in $(d+1)$-dimensional space-time $({\bf x},t)$ with the damping wave front equation given by $x_1^2+x_2^2+...+x_d^2=(R-t)^2$ and $t\ge R$. We note that the Dirac cone in this article means a complete Dirac cone or a half Dirac cone.

     (2) ${\bf iX_{eff}( {\bf k})= k_1\mathbb{I}+i(d+E_B)\mathbb{I}}$.  \\
     If $iX_{eff}( {\bf k})= k_1\mathbb{I}+i(d+E_B)\mathbb{I}$, we have $d=1$ and
     \[
     v_{eff}=\frac{\partial(Re(E_{\alpha}^{eff}( k_1)))}{\partial k_1}=1.
     \]
      Substituting this into Eq.(\ref{GF}), we get that the wave front after time $t_0$ is a point $-R+t_0$. The wave front has a Dirac cone structure in $(1+1)$-dimensional space-time $({\bf x},t)$ (damping wave front equation: $x_1=-R+t$).

    Combining cases 1 and 2, we get that the damping wave front equation has the same structure as the dispersion relation of Eq.(\ref{HDirac}) (substitute $({\bf x},t)$ with $({\bf k},E)$ in the damping wave front equation). Q.E.D.

    It is worth asking that: if a quadratic Lindbladian system has a finite system size, e.g., a $d$-dimensional disk $D^d$ ($x_1^2+x_2^2+...+x_d^2<R^2$, ${\bf x}=(x_1,x_2,...,x_d)$ is the coordinate and $R$ is the radius) which is fully filled at the initial time, whether the proposition is also true? For some 1D classes, it is true. Here we give two examples: (1) For 1D chiral edge states of a two-dimensional (2D) Chern insulator of Hermitian class A, there exits a corresponding 1D chiral damping whose damping matrix multiplying $i$ belongs to class A$^\dagger$ \cite{FSong1}. (2) For 1D helical edge states of 2D quantum spin Hall insulator of Hermitian class AII, there is a corresponding 1D helical damping whose damping matrix multiplying $i$ belongs to class AII$^\dagger$ \cite{CHLiu3}. For general dimension and classes, it is still an open question.

 Here we provide a general method to construct the quadratic Lindbladian system which has the corresponding damping modes. For the 1D chiral (helical) edge states of a 2D Chern insulators (quantum spin Hall insulators) in symmetry class A (AII), the damping matrix of 1D quadratic Lindbladian system multiplying $i$ belongs to the class A$^\dagger$ (AII$^\dagger$). It has been uncovered that the damping wave front has chiral (helical) structure \cite{FSong1,CHLiu3}.
Furthermore, in the Appendix C, we construct models with new damping modes called the 2D (3D) {\it Dirac damping} in the class DIII$^{\dagger}$ (A$^{\dagger}$). A $d$-dimensional Dirac damping  is characterized by the existence of  damping wave front having a $d$-dimensional Dirac cone structure in space-time $({\bf x},t)$. As a special case, the chiral (helical) damping is a 1D chiral (helical) Dirac damping.

\section{The value of information constraint}
       In order to describe information constraint quantitatively, we define the value of information constraint as
    \begin{equation}
   \begin{split}
   I_C(j_1,j_2,t)=&\frac{I_+(j_1,j_2,t)}{I_-(j_1,j_2,t)},
         \end{split}
    \label{ir}
    \end{equation}
     where $I_{+}$ $(I_{-})$ represents the strength of information propagating along the $+$ $(-)$ direction. It is defined as
     \begin{equation}
    \begin{split}
     I_{+}(j_1,j_2,t)=\langle \langle j_2 |e^{Xt}|j_1 \rangle \rangle \langle \langle  j_1|e^{X^{\dagger}t}|j_2  \rangle \rangle ,\\
     I_{-}(j_1,j_2,t)=\langle \langle j_1|e^{Xt}|j_2\rangle \rangle \langle \langle  j_2|e^{X^{\dagger}t}|j_1 \rangle \rangle .
     \end{split}  \label{ipm1}
    \end{equation}
     where $X$ denotes the damping matrix, $j_1=x_1q_1,j_2=x_2q_2$, $x_1,x_2\in\left\{1,2,...,L\right\}$ is the cell index, $L$ is the system size, and $q_1,q_2\in\left\{A,B\right\}$. We choose $t$ and $|x_1-x_2|\sim O(1) \ll L$ to preserve the locality of $I_{\pm}$. Here $|j_1 \rangle \rangle$ and $|j_2 \rangle \rangle$ are $2L\times1$ matrices, which are matrix representations of $|j_1\rangle$ and $|j_2\rangle$ in the single particle basis ($\left[|1A \rangle, |1B \rangle, ..., |LB \rangle \right]$), and
     $ \langle \langle j_1|$ and $ \langle \langle j_2|$ are  Hermitian conjugations of $|j_1 \rangle \rangle$ and $|j_2\rangle \rangle$.

     Corresponding to Eq.(\ref{Xk}), the damping matrix under OBC can be represented as
     \[
     X=S(-\frac{\gamma}{2}\mathbb{I}+iH_{SSH})S^{-1},
     \]
      where $H_{SSH}$ is the matrix representation of the SSH Hamiltonian under OBC with two hoping parameters  $\tilde{t}_1=\sqrt{(t_1-\frac{\gamma}{2})(t_1+\frac{\gamma}{2})}$ and $\tilde{t}_2=t_2$ in the single particle basis ($\left[|1A \rangle, |1B \rangle, ..., |LB \rangle \right]$), and
       \[
       S=diag[1,\beta,\beta,\beta^2,...,\beta^{m-1},\beta^m,...,\beta^{L-1},\beta^L]
       \]
       with
       \[
       \beta=\sqrt{\frac{t_1+\gamma/2}{t_1-\gamma/2}}.
       \]
    Substituting $j_1=x_1B$, $j_2=x_1+m B$ ($m>0$ is an integer) and $X=S(-\frac{\gamma}{2}\mathbb{I}+iH_{SSH})S^{-1}$ into Eq.(\ref{ipm1}), we get that
    \begin{eqnarray*}
    I_{+} &=& e^{2ln(\beta)m-\gamma t\mathbb{I}}| \langle \langle x_1+m B|e^{iH_{SSH}t}|x_1B \rangle \rangle|^2,\\
    I_{-} &=& e^{-2ln(\beta)m-\gamma t\mathbb{I}}| \langle \langle x_1B|e^{iH_{SSH}t}|x_1+m B \rangle \rangle|^2
    \end{eqnarray*}
    and
    \[
    I_C \approx \beta^{4m}\approx e^{1.69m}.
    \]
    For Fig.\ref{fig1}(c1), we numerically obtain
    \[
    \frac{max(\tilde{n}_{x_1+m})}{max(\tilde{n}_{x_1-m})} \approx e^{1.72m}
    \]
    under OBC, and it is approximately equal to $I_C$.
     It illustrates that $I_C$ can describe the different decay rates of signal strengths propagating along opposite directions.

    We find that $I_{\pm}$ can be alternatively defined as
      \begin{equation}
      \begin{split}
      I_{+}=|G_{j_1,j_2}(t)|^2=\left|  Tr\left[\left\{c_{j_2}(t), c^{\dagger}_{j_1}(0) \right\} \rho_{NESS}\right]\right|^2, \\
      I_{-}=|G_{j_2,j_1}(t)|^2=| Tr\left[ \left\{ c_{j_1}(t) , c^{\dagger}_{j_2}(0) \right\}\rho_{NESS} \right]|^2 \label{ipm2},
      \end{split}
      \end{equation}
      where $G_{j_1,j_2}(t)$ is the two-point Green function, and $\rho_{NESS}$ is the density matrix of the non-equilibrium steady state (NESS).  A proof of the equivalence of definitions (\ref{ipm1}) and (\ref{ipm2}) is given in Appendix D.  Here we choose $vt, ~|x_1-x_2|\sim O(1)\ll L$  (where $L$ is the system size) to preserve the locality of the Green function. The creation and annihilation operators $c^{\dagger}_{j_1}(0)$ and $c_{j_2}(t)$  satisfy the Lindblad equation in the Heisenberg picture:
     \begin{equation}
    \frac{dO}{dt} 
    =\mathcal{L}^{\dagger}[O]=i[H,O]+\sum _{\mu}\left(2L_{\mu}^{\dagger} O L_{\mu}-\left\{L_{\mu}^{\dagger}L_{\mu},O\right\}\right),
    \label{operatoreq}
    \end{equation}
    where $O$ can be any operator (for example, $c_{j_2}(t)$), and the density matrix does not evolve in this picture. $I_{\pm}$ represent the square of the absolute value of Green function. The definition of $I_C$ given by Eqs.(\ref{ir}) and (\ref{ipm1}) requires the system to be a quadratic Lindbladian system with NESS in order to make the $X$ matrix be well defined. The definition of $I_C$ given by Eqs.(\ref{ir}) and (\ref{ipm2}) only need the existence of a NESS. Thus, the definition of $I_C$ given by Eqs.(\ref{ir}) and (\ref{ipm2}) is more general than Eqs.(\ref{ir}) and (\ref{ipm1}), despite the fact that they are equivalent for some specific models.

    In the quantum viewpoint, $|G_{j_1,j_2}(t)|^2$ is the probability creating a particle at space-time $(j_1,0)$ and annihilating at $(j_2,t)$, and $G_{j_1,j_2}(t)$ contains all dynamical information of the system.
    Thus,
    \[
    I_C(j_1,j_2,t)=\frac{|G_{j_1,j_2}(t)|^2}{|G_{j_2,j_j}(t)|^2}
    \]
    can represent the ratio of decay rates of signal strengths along the $+x$ direction and $-x$ direction.

    We derive the analytical representation of Eq.(\ref{ir}) for a general $d$ dimensional quadratic Lindbladian system in the Appendix E, which is represented as
 \begin{equation}
  I_C(j_1,j_2,t)=\frac{|T(j_1,j_2,t)|^2}{|T(j_2,j_1,t)|^2}
\label{Ird-m}
\end{equation}
with
   \begin{equation}
\begin{split}
&T(j_1,j_2,t) =\langle \langle j_2 |e^{Xt}|j_1 \rangle \rangle  \\
=&\sum_{{\bf k},\alpha} \langle \langle q_2|\psi({\bf k},\alpha)\rangle \rangle_{RL}\langle \langle \psi({\bf k},\alpha)|q_1 \rangle \rangle e^{E_{\alpha}({\bf k})t+i{\bf k}({\bf x}_2-{\bf x}_1)},
\end{split} \label{Tpbcd-m}
\end{equation}
 where $X({\bf k})$ is the damping matrix in momentum space,  $j_1={\bf x}_1q_1$, $j_2={\bf x}_2q_2$, the $d$-dimensional vectors ${\bf x}_1$ and ${\bf x}_2$ label the location of cells, $q_1$ and $q_2$ label the degree of freedom in the cell. Here ${\bf k}$ is the $d$-dimensional momentum, $\alpha$ is the band index of $X({\bf k})$, and $E_{\alpha}({\bf k})$, $|{\bf k},\alpha\rangle \rangle_R$ and $|{\bf k},\alpha\rangle \rangle_L$ are the eigenvalues, right eigenvectors and left eigenvectors of $X({\bf k})$, respectively. We denote $|{\bf k} ,\alpha\rangle \rangle_R=|{\bf k}\rangle \rangle \otimes |\psi({\bf k},\alpha)\rangle \rangle_R$, $_ L\langle \langle {\bf k} ,\alpha|=\langle \langle {\bf k}| \otimes$$_L\langle \langle\psi({\bf k},\alpha)|$, $|x_1q_1\rangle \rangle=|{\bf x}_1\rangle \rangle \otimes |q_1\rangle \rangle$ and $|{\bf x}_2q_2\rangle \rangle=|{\bf x}_2\rangle \rangle \otimes |q_2\rangle \rangle$, where $|\psi({\bf k},\alpha)\rangle \rangle_R$, $|\psi({\bf k},\alpha)\rangle \rangle_L$, $|q_2\rangle \rangle$ and $|q_2\rangle \rangle$ belong to the Hilbert space in the unit cell, and $|k\rangle \rangle$, $|{\bf x}_1\rangle \rangle$ and $|{\bf x}_2\rangle \rangle$ belong to the Hilbert space of cell index. In the Appendix E, we use Eqs.(\ref{Ird-m}) and (\ref{Tpbcd-m}) to calculate $I_C$ for the model described by Eqs.(\ref{operators2}) and (\ref{operators1}), and get $I_C\approx e^{1.6m}$, which is consistent with our pervious result $I_C\approx e^{1.69m}$ obtained under OBC. Here we note that  the result $I_C\approx e^{1.6m}$ under PBC is obtained analytically after taking some approximations. A more accurate numerical calculation gives that $I_C\approx e^{1.69m}$ even under PBC.
  We also give an analytical derivation of chiral damping and helical damping via information constraint in the Appendix F.

      The value $I_C=1$ means the vanishing of information constraint. For a quadratic Lindbladian system, if the damping matrix $X$ satisfies that $X^{T}=X$, then $I_C=1$. The proof is given in the Appendix G. In general, if there is no symmetry constraint, $I_C\ne 1$.

\section{Summary and discussion}
In summary, we propose an effect coined information constraint which is an intrinsic
 property of a type of open quantum systems independent of the boundary condition. We define the value of information constraint $I_C$ and illustrate that it can effectively describe the ratio of different decay rates of signal strengths propagating along opposite directions. We derive the analytical representation of $I_C$ for general quadratic Lindbladian systems. Based on information constraint, we can get a simple and elegant explanation for the chiral and helical damping, and also get the local maximum points of $ln|\tilde n_x(t)|$ of the periodic system, which is consistent with the numerical calculation. The model with the helical damping is predicted to have the helical tunneling effect.
Inspired by information constraint, we propose and prove the correspondence between $d$-dimensional anomalous edge modes of $(d+1)$-dimensional close quantum system and $d$-dimensional damping modes of quadratic Lindbladian systems. A new damping mode called Dirac damping is constructed.

\begin{acknowledgments}
C.-H. Liu  would thank K. Zhang, Z. Yang and Z. Wang for very helpful discussions. The work is supported by the  NSFC under Grants No.11974413 and the Strategic Priority Research Program of Chinese Academy of Sciences under Grant No. XDB33000000.
\end{acknowledgments}

\appendix
\section{Derivation of tunneling amplitude for the helical damping model and demonstration of helical tunneling behavior.}
Consider the model discussed in Ref.\cite{CHLiu3}. For convenience, here we write this model explicitly with the Hamiltonian described by
    \begin{equation}
    h(k)=t_1\sigma_x+(t_2\sigma_y+\delta_1\tau_x) \sin k+t_2\sigma_x \cos k, \label{h}
    \end{equation}
    and the Lindblad operators
   \begin{equation}
   \begin{split}
    L_{x\uparrow}^l=\sqrt{\frac{\gamma_l}{2}}(c_{xA\uparrow}-ic_{xB\uparrow}), \quad L_{x\uparrow}^g=\sqrt{\frac{\gamma_g}{2}}(c^{\dagger}_{xA\uparrow}+ic^{\dagger}_{xB\uparrow}), \\
    L_{x\downarrow}^l=\sqrt{\frac{\gamma_l}{2}}(c_{xA\downarrow}+ic_{xB\downarrow}), \quad L_{x\downarrow}^g=\sqrt{\frac{\gamma_g}{2}}(c^{\dagger}_{xA\downarrow}-ic^{\dagger}_{xB\downarrow}) .
    \end{split}
    \label{operators}
    \end{equation}
  Here $A,B$ represent the orbit and $\uparrow,\downarrow$ represent the spin, $\sigma_x$, $\sigma_y$, $\sigma_z$ act on orbit degree of freedom, and $\tau_x$, $\tau_y$, $\tau_z$ act on spin degree of freedom. The damping matrix is
  \begin{eqnarray}
        X &=& i\left[ \begin{array}{cc}
            H_{nSSH}(k) +\frac{i\gamma}{2}& \delta_1 \sin k \\
            \delta_1 \sin k& H_{nSSH}^T(-k)+\frac{i\gamma}{2}
            \end{array}
            \right ] \nonumber \\
             &=&(-\frac{\gamma}{2}+it_1\sigma_x+\frac{\gamma}{2}\sigma_y\tau_z)+i(t_2\sigma_y+\delta_1\tau_x) \sin k \nonumber\\
             & &+it_2\sigma_x \cos k,  \label{xmatrix}
        \end{eqnarray}
  where $\gamma=\gamma_l+\gamma_g$ and
  \[
  H_{nSSH}(k)=(t_1+t_2 \cos k)\sigma_x+(t_2 \sin k-\frac{i\gamma}{2})\sigma_y.
  \]
  It fulfills
        \begin{equation}
        CX(-k)^T=X(k)C \label{trsymmetry}
        \end{equation}
        with $C=i\tau_y$.

Next we define $T_{(x_1,s_1,o_1)\rightarrow(x,s,o)}$ and $ T_{(x_1,s_1,o_1)\rightarrow(x,s,o),k+i\kappa}$ as
\begin{equation}
\begin{split}
T_{(x_1,s_1,o_1)\rightarrow(x,s,o)}=&\langle \langle(x,s,o)|e^{Xt}|(x_1,s_1,o_1)\rangle\rangle
\end{split}  \label{Txso}
\end{equation}
and
\begin{equation}
\begin{split}
T_{(x_1,s_1,o_1)\rightarrow(x,s,o),k+i\kappa}=&\langle \langle(x,s,o)|e^{X(k+i\kappa)t}|(x_1,s_1,o_1)\rangle\rangle
\end{split}, \label{Txsok}
\end{equation}
where $T_{(x_1,s_1,o_1)\rightarrow(x,s,o)}$ is the tunneling amplitude from $(x_1,s_1,o_1)$ to $(x,s,o)$ and $T_{(x_1,s_1,o_1)\rightarrow(x,s,o),k+i\kappa}$ is the $k+i\kappa$ component of this tunneling amplitude.  Here all $z=e^{k+i\kappa}$ constitute the GBZ of damping matrix $X$, $\kappa$ is  a function of $k$ and $\alpha$, and $\alpha$ denotes the band index of $X(k+i\kappa)$ \cite{CHLiu3}. For convenience, we use $\kappa$ to represent $\kappa(k,\alpha)$, $|(x,s,o)\rangle\rangle$ and $|(x_1,s_1,o_1)\rangle\rangle$ to denote matrix representation of $|(x,s,o)\rangle$ and $|(x_1,s_1,o_1)\rangle$ in the single particle basis $[(1,\uparrow,A),(1,\uparrow,B),(1,\downarrow,A),(1,\downarrow,B),...,(L,\uparrow,A),(L,\uparrow,B),(L,\downarrow,A),(L,\downarrow,B)]$, and $s\in\left\{\uparrow,\downarrow\right\}$ and $o\in\left\{A,B\right\}$ to represent the spin and orbit degree of freedom, respectively.  According to non-Bloch band theory, $T_{(x_1,s_1,o_1)\rightarrow(x,s,o)}=\sum_{k,\alpha} T_{(x_1,s_1,o_1)\rightarrow(x,s,o),k+i\kappa}$. Before deducing the formula of $T_{(x_1,s_1,o_1)\rightarrow(x,s,o)}$, some notions or formulas should be introduced: $E(k+i\kappa)$, $|k+i\kappa,\alpha \rangle \rangle_R$ and $|k+i\kappa,\alpha\rangle \rangle_L$ are the eigenvalues, right eigenvectors and left eigenvectors of $X(k+i\kappa)$, respectively, where
  \begin{eqnarray*}
  &|k+i\kappa ,\alpha\rangle \rangle_R=|k+i\kappa\rangle \rangle_R \otimes |\psi(k+i\kappa,\alpha)\rangle \rangle_R, \\
  & _L\langle \langle k+i\kappa,\alpha|=_L\langle \langle k+i\kappa| \otimes_L\langle \langle\psi(k+i\kappa,\alpha)|
 \end{eqnarray*}
   and
   \[
   |(x,s,o)\rangle \rangle=|x\rangle \rangle \otimes |(s,o)\rangle \rangle.
   \]
    Here $|\psi(k+i\kappa,\alpha)\rangle \rangle_R$, $|\psi(k+i\kappa,\alpha)\rangle \rangle_L$ and $|(s,o)\rangle \rangle$ belong to the Hilbert space inside the unit cell, and $|k+i\kappa\rangle \rangle_R$, $|k+i\kappa\rangle \rangle_L$ and $|x\rangle \rangle$ belong to the Hilbert space of cell index. And we have
   \begin{eqnarray*}
   &~~~~ & \langle \langle x|k+i\kappa\rangle \rangle_R=e^{i(k+i\kappa)x}, \\
   &~~~~&   \langle \langle x|k+i\kappa\rangle \rangle_L=e^{i(k-i\kappa)x}, \\
   & & \langle \langle (x,s,o)|k+i\kappa,\alpha\rangle \rangle_R \\
    &=& \langle \langle x| k+i\kappa \rangle \rangle_R \langle \langle (s,o))| \psi(k+i\kappa,\alpha) \rangle \rangle_R \\
  &=& e^{i(k+i\kappa)x}\langle \langle (s,o)| \psi(k+i\kappa,\alpha) \rangle \rangle_R
  \end{eqnarray*}
   and
   \[
   \sum_{k',\alpha'}|k'+i\kappa',\alpha'\rangle \rangle_{RL}\langle \langle k'+i\kappa',\alpha'|=\mathbb{I}.
   \]
   Taking these into account, from Eq.(\ref{Txso}), we have
\begin{equation}
\begin{split}
&T_{(x_1,s_1,o_1)\rightarrow(x,s,o)}\\
=&\sum_{k,\alpha} T_{(x_1,s_1,o_1)\rightarrow(x,s,o),k+i\kappa} \\
=&\sum_{k,\alpha} \langle\langle(x,s,o)|e^{X(k+i\kappa)t}|(x_1,s_1,o_1)\rangle \rangle \\
=&\sum_{k,\alpha,k',\alpha'}\langle\langle(x,s,o)|e^{X(k+i\kappa)t}|k'+i\kappa',\alpha\rangle\rangle_{R}\times\\
& _{L}\langle \langle k'+i\kappa',\alpha|(x_1,s_1,o_1)\rangle\rangle\\
=&\sum_{k,\alpha}\langle\langle (s,o)|\psi(k+i\kappa,\alpha)\rangle\rangle_{R}\times \\
&_{L}\langle \langle \psi(k+i\kappa,\alpha)|(s_1,o_1)\rangle\rangle e^{i(k+i\kappa)(x-x_1)+E(k+i\kappa)t} .
\end{split}
\end{equation}

  Consider the case that the real part of the continuous spectrum of $X$ under OBC approximately equals to  $\gamma/2$ (it can be represented as $Re(E(k+i\kappa))\approx \gamma/2$). In fact, if $\gamma \ll t_1,t_2$ or $t_1=t_2=1,\gamma=0.4,\delta_1=0.1$ or $t_1=1.2, t_2=1, \gamma=0.2, \delta_1=0.1$, we can get that $Re(E(k+i\kappa))\approx \gamma/2$\cite{CHLiu3}. Thus, the three situations all fall into this case.

 Assume that we have $\kappa=maximum(\kappa)=\kappa_{max}$ at point $k=k_{1}$ and $\alpha=\alpha_{1}$, and $\kappa=minimum(\kappa)=\kappa_{min}$ at point $k=k_{2}$ and $\alpha=\alpha_{2}$. Because of the symmetry of Eq.(\ref{trsymmetry}), the spin-orbit components $\alpha_{1}$ and $\alpha_{2}$ have opposite spins and  correspondingly $\kappa_{max}=-\kappa_{min}=\kappa_0>0$. Together with $Re(E(k+i\kappa))\approx \frac{\gamma}{2}$, if $x_1>x$,  $|T_{(x_1,s_1,o_1)\rightarrow(x,s,o)}|$ is dominated by $k=k_{1}$ and $\alpha=\alpha_{1}$ component:
\begin{equation}
\begin{split}
&|T_{(x_1,s_1,o_1)\rightarrow(x,s,o)}||_{x_1>x}\\
&\approx |\langle \langle(s,o)|\psi(k+i\kappa,\alpha)\rangle\rangle_{R}\times\\
&_{L}\langle \langle \psi(k+i\kappa,\alpha)|(s_1,o_1)\rangle \rangle||_{k=k_1,\alpha=\alpha_1} e^{\kappa_{max}(x_1-x)-\frac{\gamma}{2}t},  \\
=&  |\langle \langle(s,o)|\psi(k+i\kappa,\alpha)\rangle\rangle_{R}\times \\
&_{L}\langle \langle \psi(k+i\kappa,\alpha)|(s_1,o_1)\rangle \rangle||_{k=k_1,\alpha=\alpha_1} e^{\kappa_0(x_1-x)-\frac{\gamma}{2}t}.
\end{split}\label{Tx12a}
\end{equation}
If $x>x_1$,  $|T_{(x_1,s_1,o_1)\rightarrow(x,s,o)}|$ is dominated by $k=k_{2}$ and $\alpha=\alpha_{2}$ component:
\begin{equation}
\begin{split}
&|T_{(x_1,s_1,o_1)\rightarrow(x,s,o)}||_{x_1<x} \\
&\approx  |\langle \langle(s,o)|\psi(k+i\kappa,\alpha)\rangle\rangle_{R}\times\\
&_{L}\langle \langle \psi(k+i\kappa,\alpha)|(s_1,o_1)\rangle \rangle||_{k=k_2,\alpha=\alpha_2} e^{\kappa_{min}(x_1-x)-\frac{\gamma}{2}t},  \\
&=|\langle \langle(s,o)|\psi(k+i\kappa,\alpha)\rangle\rangle_{R}\times\\
&_{L}\langle \langle \psi(k+i\kappa,\alpha)|(s_1,o_1)\rangle \rangle||_{k=k_2,\alpha=\alpha_2} e^{-\kappa_0(x_1-x)-\frac{\gamma}{2}t}.
\end{split}  \label{Tx12i}
\end{equation}
Eq.(\ref{Tx12a}) shows that the $\alpha_1$ component tends to tunneling through the ``$-$" direction (from $x_1$ to $x$ and $x_1>x$). Eq.(\ref{Tx12i}) shows that the $\alpha_2$ component tends to tunneling through the ``$+$" direction (from $x_1$ to $x$ and $x_1<x$). Since the spin-orbit components $\alpha_{1}$ and $\alpha_{2}$ have opposite spins, the model shows a helical tunneling behavior.

\section{Open Heisenberg XX spin chain with information constraint}
The information constraint also exists in the Heisenberg XX spin chain. Consider the Lindbald master equation with the Hamiltonian described by
    \begin{equation}
   \begin{split}
    h=&\sum_{j=1}^{N} J_1(\sigma_{2j-1}^x\sigma_{2j}^x+\sigma_{2j-1}^y\sigma_{2j}^y) \\
    +&\sum_{j=1}^{N-1}J_2(\sigma_{2j}^x\sigma_{2j+1}^x+ \sigma_{2j}^y\sigma_{2j+1}^y)
    \end{split}
    \label{interaction1}
 \end{equation}
and the Lindblad operators given by
   \begin{equation}
   \begin{split}
    L_{j}=\sqrt{g}(\sigma_{2j-1}^--i\sigma_{2j}^-). \\
    \end{split}
    \label{interaction2}
    \end{equation}
     If we omit the quantum jump term in the master equation, the evolution of density matrix is governed by
     \[
     \rho(t)=e^{-iH_{NH}t}\rho(0)e^{iH_{NH}^{\dagger}t}
     \]
     with $H_{NH}=h-i\sum_j L_j^{\dagger}L_j$. After the Jordan-Wigner transformation, we get
      \begin{equation}
   \begin{split}
   H_{NH}=&\sum_{j=1}^{2N}-ig a_j^{\dagger}a_j+\sum_{j=1}^N[(2J_1-g)a_{2j-1}^{\dagger}a_{2j}+\\
   &(2J_1+g)a_{2j}^{\dagger}a_{2j-1}]+
     \sum_{j=1}^{N-1}[2J_2a_{2j}^{\dagger}a_{2j+1}+ h.c.].
         \end{split}
    \label{interaction3}
    \end{equation}
   Here $H_{NH}$ plays a similar role as the damping matrix $X$ in the main text (If we expand $H_{NH}$ in the invariant subspace spanned by $|1 \rangle, |2 \rangle, ..., |2N \rangle$, $H_{NH}$ and $X$ have the same formula). It has been shown that the model in the main text has information constraint, and thus we can get that information constraint exists in the open Heisenberg XX spin chain.

\section{Correspondence between damping modes and edge modes}

The section includes two subsections. In the first subsection, we introduce the Hermitian and non-Hermitian symmetry class. In the second subsection, we present some examples of the proposition II.

\subsection{Hermitian and non-Hermitian symmetry class}
          \begin{table}[h]
        \caption{AZ class. $U_TU_T^*=0$, $U_PU_P^*=0$ and $U_S^2=0$ represent that there is no $T$, $P$ and $S$ symmetry, respectively.}
      \begin{center}
      \begin{tabular}{cccccc}
       \hline\hline
       s&\; AZ class\;&$U_TU_T^*$ &$U_PU_P^*$ & $ U_S^2$&Classifying Space\\
        \hline\hline
        Complex case\\
        0&A&$0$ & $0$ & $0$&$C_0$\\
        1&AIII&$0$ & $0$ & $1$&$C_1$\\
        Real case\\
        0&AI&$1$ & $0$ & $0$ &$R_0$\\
        1&BDI&$1$ & $1$ &$1$ &$R_1$\\
        2&D&$0$ & $1$ & $0$ &$R_2$\\
        3&DIII&$-1$ & $1$ & $1$&$R_3$\\
        4&AII&$-1$ & $0$ & $0$&$R_4$\\
        5&CII&$-1$ & $-1$ & $1$&$R_5$\\
        6&C&$0$ & $-1$ & $0$ &$R_6$\\
        7&CI&$1$ & $-1$ & $1$&$R_7$\\
      \hline\hline
        \end{tabular}
      \end{center} \label{AZclass}
    \end{table}
 {\bf Altland-Zirnbauer class.} The Hermitian system is described by Altland-Zirnbauer (AZ) class. There are three types of symmetries: time-reversal symmetry ($T$), particle-hole symmetry ($P$) and sublattice symmetry ($S$) which fulfill that,
   \begin{align}
   U_{T}H^*(-{\bf k})U_{T}^{-1}=H({\bf k}) &, U_{T}U_{T}^*=\eta_T \mathbb{I}&T \textrm{ sym.} \label{eq:syms1} \\
   U_{P}H^*(-{\bf k})U_{P}^{-1}=-H({\bf k}) &, U_{P}U_{P}^*=\eta_P \mathbb{I}&P \textrm{ sym.} \label{eq:syms1}\\
   U_{S}H({\bf k})U_{S}^{-1}=-H({\bf k})  &, U_{S}^2=\mathbb{I} &S \textrm{ sym.} \label{AZsym}
      \end{align}
 where $\eta_T,\eta_P=\pm 1$ and $S=TP$. These symmetries can constitute tenfold AZ classes. The tenfold AZ classes include two complex classes ($s=1,2$) and eight real classes ($s=1,2,...,8$) which are shown in Table \ref{AZclass}.

{\bf Bernard-LeClair class.}  The non-Hermitian system is described by 38-fold Bernard-LeClair (BL) classes for point gap systems \cite{Sato,Zhou} and 54-fold generalized Bernard-LeClair (GBL) classes for line gap systems \cite{CHLiu2}. There are four types of symmetries: P, Q, C and K, which fulfill that,
\begin{align}
    H({\bf k})=\epsilon_kkH({\bf k})^*k^{-1}&, ~~kk^*=\eta_k \mathbb{I}, &K \textrm{ sym.} \label{eq:syms1} \\
    H({\bf k})=\epsilon_qqH({\bf k})^\dagger q^{-1}&, ~~q^2=\mathbb{I}, &Q \textrm{ sym.}\label{eq:syms2}\\
    H({\bf k})=\epsilon_c cH({\bf k})^Tc^{-1}&, ~~cc^*=\eta_c \mathbb{I},&C \textrm{ sym.}\label{eq:syms3}\\
    H({\bf k})=-pH({\bf k})p^{-1}&, ~~p^2=\mathbb{I}, &P \textrm{ sym.}\label{GBLsym}
    \end{align}
with
    \begin{equation}
        c=\epsilon_{pc}pcp^T, \quad k=\epsilon_{pk}pkp^T, \quad c=\epsilon_{qc}qcq^T, \quad p=\epsilon_{pq}qpq^\dagger.
      \end{equation}
    For point gap systems, $H\rightarrow iH$ is an equivalent transformation. Due to $\epsilon_k, \epsilon_q=1$ and $\eta_k,\epsilon_c,\eta_c,\epsilon_{pc},\epsilon_{pk},\epsilon_{qc}$, $\epsilon_{pq}=\pm1$, these symmetries can constitute 38-fold BL classes \cite{Sato,Zhou}.  For line gap systems, $H\rightarrow iH$ is not an equivalent transformation. Due to $\epsilon_k,\epsilon_q,\eta_k,\epsilon_c,\eta_c,\epsilon_{pc},\epsilon_{pk},\epsilon_{qc}$, $\epsilon_{pq}=\pm1$, these symmetries can constitute 54-fold GBL class\cite{CHLiu2}.

    {\bf AZ$^{\dagger}$ class.} AZ$^{\dagger}$ class is a subset of the BL or GBL class. If we substitute the time-reversal symmetry with $C$ symmetry ($\epsilon_c=1,\eta_c=\pm1$), the particle-hole symmetry with $K$ symmetry ($\epsilon_k=-1,\eta_k=\pm1$) and the sublattice symmetry with $Q$ symmetry ($\epsilon_q=-1$) in the AZ class, we can get AZ$^{\dagger}$ class \cite{Sato}. Three types of symmetries of AZ$^{\dagger}$ classes,  which fulfill that
\begin{align}
  cH^T(-{\bf k})c^{-1}=H({\bf k}) &, cc^*=\eta_c \mathbb{I} \\
   kH^*(-{\bf k})k^{-1}=-H({\bf k}) &, kk^*=\eta_k \mathbb{I}\\
   qH^{\dagger}({\bf k})q^{-1}=-H({\bf k})  &, q^2=\mathbb{I}  \label{AZdsym}
      \end{align}
      where $\eta_c,\eta_k=\pm 1$. These symmetries can constitute 10-fold AZ$^{\dagger}$ classes. The 10-fold AZ$^{\dagger}$ classes include two complex classes ($s=1,2$) and eight real classes ($s=1,2,...,8$) which are shown in Table \ref{AZdclass}.

         \begin{table}[h]
        \caption{AZ$^{\dagger}$ class. $cc^*=0$, $kk^*=0$ and $q^2=0$ represent that there is no $C$, $K$ and $Q$ symmetry, respectively.}
      \begin{center}
      \begin{tabular}{cccccc}
       \hline\hline
       s$^{\dagger}$&\;\;\; AZ$^{\dagger}$ class\;\;\;&$cc^*$ &$kk^*$ & $ q^2$\\
        \hline\hline
        Complex case\\
        0$^{\dagger}$&A$^{\dagger}$&$0$ & $0$ & $0$\\
        1$^{\dagger}$&AIII$^{\dagger}$&$0$ & $0$ & $1$\\
        Real case\\
        0$^{\dagger}$&AI$^{\dagger}$&$1$ & $0$ & $0$ \\
        1$^{\dagger}$&BDI$^{\dagger}$&$1$ & $1$ &$1$ \\
        2$^{\dagger}$&D$^{\dagger}$&$0$ & $1$ & $0$ \\
        3$^{\dagger}$&DIII$^{\dagger}$&$-1$ & $1$ & $1$\\
        4$^{\dagger}$&AII$^{\dagger}$&$-1$ & $0$ & $0$\\
        5$^{\dagger}$&CII$^{\dagger}$&$-1$ & $-1$ & $1$\\
        6$^{\dagger}$&C$^{\dagger}$&$0$ & $-1$ & $0$ \\
        7$^{\dagger}$&CI$^{\dagger}$&$1$ & $-1$ & $1$\\
      \hline\hline
        \end{tabular}
      \end{center} \label{AZdclass}
    \end{table}

    \subsection{Examples}
  In this section, we discuss some examples which have the corresponding damping modes.

    {\bf 1D A$^{\dagger}$ class.} According to the classification, there exits a 1D surface half Dirac cone for the topologically non-trivial 2D Hermitian A class. The 1D surface Dirac cone is characterized by
  \begin{equation}
   \begin{split}
  H_{1DA}(k)=k.
  \end{split}  \label{H1DA}
    \end{equation}
     According to Eq.(\ref{H1DA}), we construct the damping matrix as,
     \begin{equation}
   \begin{split}
  X_{1DA}(k)=-i \sin(k)+[\cos(k)-1].
  \end{split}  \label{X1DA}
    \end{equation}

     We can verify that $iX_{1DA}$ belongs to class A$^{\dagger}$. According to Eq.(\ref{X1DA}), we construct the quadratic Lindbladian system described by
    \begin{equation}
    h(k)=\sin(k),
    \end{equation}
   \begin{equation}
   \begin{split}
    L_{x}=\frac{1}{\sqrt{2}}(-c_{x}+c_{x+1}),
    \end{split}
    \end{equation}
    where $c_{x}$ is an annihilate operator at cell $x$. Consider this model with the infinite boundary condition, and it is fully filled in a 1D disk $D^1$ ( $0<x<R$, where $R$ is constant) and empty in $x<0$ and $x>R$. To get the longest life-time effective theory, we expand $iX_{1DA}(k)$ at $ k=0$, which gives rise to
     \begin{equation}
   \begin{split}
  iX_{1DA}^{eff}(k)=k.
  \end{split}  \label{X1DAeff}
    \end{equation}
It follows
     $v_{eff}=\frac{\partial Re(E_{\alpha}^{eff}(k))}{\partial k}=1$, where $E_{\alpha}^{eff}$ is the eigenvalue of $iX_{1DA}^{eff}(k)$.

    Substituting Eq.(\ref{X1DAeff}) into Eq.(\ref{GF}) and focusing on the points in the $D^1$, we get the wave front after time $t_0$ located at $t_0$. The wave front has a half Dirac cone structure in $1+1$ dimensional space-time $(x,t)$ (damping wave front equation: $x=t$). The damping behavior is chiral damping.

        {\bf 1D DIII$^{\dagger}$ class.} According to the classification, there is a 1D surface Dirac cone for the topologically non-trivial 2D Hermitian DIII class. The 1D surface Dirac cone is characterized by
  \begin{equation}
   \begin{split}
  H_{1DIII}(k)=k\sigma_x.
  \end{split}  \label{H1DIII}
    \end{equation}
     It fulfills
     \[
     \sigma_xH_{1DIII}^*(-k)\sigma_x=-H_{1DIII}(k)
     \]
     and
     \[
     i\sigma_yH_{1DIII}^*(-k)(-i\sigma_y)=H_{1DIII}(k).
     \]
     According to Eq.(\ref{H1DIII}), we construct the damping matrix as
     \begin{equation}
   \begin{split}
  X_{1DIII}(k)=-i \sin(k)\sigma_x+[\cos(k)-1]\sigma_0.
  \end{split}  \label{X1DIII}
    \end{equation}

    We can verify that $iX_{1DIII}$ belongs to the class DIII$^{\dagger}$, i.e.,
    \[
    \sigma_x[iX_{1DIII}(-k)]^*\sigma_x=-iX_{1DIII}(k)
    \]
    and
    \[
    i\sigma_y [iX_{1DIII}(-k)]^T(-i\sigma_y)=iX_{1DIII}(k).
    \]
    According to Eq.(\ref{X1DIII}), we construct the quadratic Lindbladian system as
    \begin{equation}
    h(k)= \sin(k)\sigma_x,
    \end{equation}
   \begin{equation}
   \begin{split}
    L_{x1}=\frac{1}{2}(c_{x,\uparrow}+c_{x,\downarrow}-c_{x+1,\uparrow}-c_{x+1,\downarrow}), \\
    L_{x2}=\frac{1}{2}(c_{x,\uparrow}-c_{x,\downarrow}-c_{x+1,\uparrow}+c_{x+1,\downarrow}),
    \end{split}
    \end{equation}
    where $c_{x,\uparrow}$ ($c_{x,\downarrow}$) is an annihilation operator at the cell $x$ for spin $\uparrow$ ($\downarrow$). Consider this model with the infinite system size,
     and it is fully filled in a 1D disk $D^1$ ($-R<x<R$, where $R$ is radius) and empty in $|x|>R$. To get the longest life-time effective theory, we expand $X_{1DIII}$ at $ k=0$, which gives rise to
     \begin{equation}
   \begin{split}
  iX_{1DIII}^{eff}(k)=k\sigma_x.
  \end{split}  \label{X1DIIIeff}
    \end{equation}
It follows
     $v_{eff}=\frac{\partial Re(E_{\alpha}^{eff}(k))}{\partial k}=\pm 1$, where $E_{\alpha}^{eff}$ is the eigenvalues of $iX_{1DIII}^{eff}(k)$.

    Substituting Eq.(\ref{X1DIIIeff}) into Eq.(\ref{GF}) and focusing on the points in the $D^1$,  we get that the wave front after time $t_0$ is a sphere $S^0$ with radius $|R-t_0|$ and center at $0$. The wave front has a Dirac cone structure in the $1+1$ dimensional space-time $(x,t)$ (damping wave front equation: $x=|R-t|$ and $t\ge R$). The damping behavior is helical damping.

  {\bf 2D DIII$^\dagger$ class.} According to the classification, there is a surface Dirac cone for the topologically non-trivial 2D Hermitian DIII class. The surface Dirac cone is characterized by
  \begin{equation}
   \begin{split}
  H_{2D}({\bf k})=k_x\sigma_x+k_y\sigma_y,
  \end{split}  \label{H2D}
    \end{equation}
    where ${\bf k}=(k_x,k_y)$. It fulfills $\sigma_zH_{2D}({\bf k})\sigma_z=-H_{2D}({\bf k})$ and $i\sigma_yH_{2D}^*(-{\bf k})(-i\sigma_y)=H_{2D}({\bf k})$. According to Eq.(\ref{H2D}), we construct the damping matrix as
     \begin{equation}
   \begin{split}
  X_{2D}({\bf k})=&-i[\sin(k_x)\sigma_x + \sin(k_y)\sigma_y] \\
  &+[\cos(k_x)+ \cos(k_y)-2] \sigma_0.
  \end{split}  \label{X2D}
    \end{equation}

    We can verify that $iX_{2D}$ belongs to the class DIII$^{\dagger}$ as it fulfills
    \[
    \sigma_z [iX_{2D}({\bf k})]^{\dagger}\sigma_z=-iX_{2D}({\bf k})
    \] and
    \[
    \sigma_y[iX_{2D}(-{\bf k})]^{T}(-i\sigma_y)=iX_{2D}({\bf k}).
    \]
    According to Eq.(\ref{X2D}), we construct the quadratic Lindbladian system as
    \begin{equation}
    h({\bf k})=\sin(k_x)\sigma_x+ \sin(k_y)\sigma_y,
    \end{equation}
   \begin{equation}
   \begin{split}
    L_{(x,y)1}=\frac{1}{2}(c_{(x,y)A}+c_{(x,y)B}-c_{(x+1,y)A}-c_{(x+1,y)B}), \\ L_{(x,y)2}=\frac{1}{2}(c_{(x,y)A}+c_{(x,y)B}-c_{(x,y+1)A}-c_{(x,y+1)B}) ,\\
     L_{(x,y)3}=\frac{1}{2}(c_{(x,y)A}-c_{(x,y)B}-c_{(x+1,y)A}+c_{(x+1,y)B}), \\ L_{(x,y)4}=\frac{1}{2}(c_{(x,y)A}-c_{(x,y)B}-c_{(x,y+1)A}+c_{(x,y+1)B}) ,
    \end{split}
    \end{equation}
    where $c_{(x,y)A}$ $(c_{(x,y)B})$ is an annihilation operator at the cell $(x,y)$ and sublattice $A$ $(B)$. Consider this model with the infinite system size, and it is fully filled in a 2D disk $D^2$ ($x^2+y^2<R^2$, where ${\bf x}=(x,y)$ is the coordinate and $R$ is the radius) and empty in $x^2+y^2>R^2$. To get the longest life-time effective theory, we expand $X_{2D}$ at ${\bf k}=(0,0)$, which gives rise to
     \begin{equation}
   \begin{split}
  iX_{2D}^{eff}({\bf k})=k_x\sigma_x+k_y\sigma_y.
  \end{split}  \label{X2Deff}
    \end{equation}
It follows
     \begin{eqnarray*}
     v_{eff}&=&\left(\frac{\partial Re(E_{\alpha}^{eff}({\bf k}))}{\partial k_x},\frac{\partial Re(E_{\alpha}^{eff}({\bf k}))}{\partial k_y}\right) \\
     &=& \pm\left(\frac{k_x}{\sqrt{k_x^2+k_y^2}},\frac{k_y}{\sqrt{k_x^2+k_y^2}}\right),
     \end{eqnarray*}
     $|v_{eff}|=1$ and $E_{\alpha}^{eff}$ is the eigenvalues of $iX_{2D}^{eff}({\bf k})$.

    Substituting Eq.(\ref{X2Deff}) into Eq.(\ref{GF}) and focusing on the points in the $D^2$,  we get that the damping wave front after time $t_0$ is a sphere $S^1$ with radius $|R-t_0|$ and center at $(0,0)$. The damping wave front has a Dirac cone structure in the $(d+1)$-dimensional space-time $({\bf x},t)$ (damping wave front equation: $x^2+y^2=(R-t)^2$ and $t\ge R$). We dub this damping behavior as a 2D Dirac damping since the damping wave front has a Dirac cone structure.

      {\bf 3D A$^\dagger$ class.} According to the classification, there is a surface Dirac cone for the topologically non-trivial 3D Hermitian A class. The surface Dirac cone is characterized by
  \begin{equation}
   \begin{split}
  H_{3D}({\bf k})=k_x \sigma_x+k_y \sigma_y+k_z \sigma_z,
  \end{split}  \label{H3D}
    \end{equation}
  where ${\bf k}=(k_x,k_y,k_z)$. According to Eq.(\ref{H3D}), we construct the damping matrix as
     \begin{equation}
   \begin{split}
  X_{3D}({\bf k})=&-i [ \sin(k_x)\sigma_x+ \sin(k_y)\sigma_y+ \sin(k_z)\sigma_z]\\
  &+ [ \cos(k_x)+ \cos(k_y)+ \cos(k_z)-3]\sigma_0.
  \end{split}  \label{X3D}
    \end{equation}
  We can verify that $iX_{3D}$ belongs to the class A$^{\dagger}$. According to Eq.(\ref{X3D}), we construct the quadratic Lindbladian system as
    \begin{equation}
    h({\bf k})= \sin(k_x) \sigma_x+ \sin(k_y)\sigma_y+ \sin(k_z)\sigma_z,
    \end{equation}
    and
   \begin{equation}
   \begin{split}
    & L_{(x,y,z)1} =\frac{1}{2}(c_{(x,y,z)A}+c_{(x,y,z)B}-c_{(x+1,y,z)A}-c_{(x+1,y,z)B}),\\
    & L_{(x,y,z)2}  =\frac{1}{2}(c_{(x,y,z)A}+c_{(x,y,z)B}-c_{(x,y+1,z)A}-c_{(x,y+1,z)B}),\\
   & L_{(x,y,z)3} =\frac{1}{2}(c_{(x,y,z)A}+c_{(x,y,z)B}-c_{(x,y,z+1)A}-c_{(x,y,z+1)B}) ,\\
    & L_{(x,y,z)4}=\frac{1}{2}(c_{(x,y,z)A}-c_{(x,y,z)B}-c_{(x+1,y,z)A}+c_{(x+1,y,z)B}),\\
    & L_{(x,y,z)5} = \frac{1}{2}(c_{(x,y,z)A}-c_{(x,y,z)B}-c_{(x,y+1,z)A}+c_{(x,y+1,z)B}),\\
   & L_{(x,y,z)6} = \frac{1}{2}(c_{(x,y,z)A}-c_{(x,y,z)B}-c_{(x,y,z+1)A}+c_{(x,y,z+1)B}) ,
    \end{split}
    \end{equation}
    where $c_{(x,y,z)A}$ $(c_{(x,y,z)B})$ is an annihilation operator at the cell $(x,y,z)$ and sublattice $A$ $(B)$. Consider this model with the infinite system size, and it is fully filled in a 3D disk $D^3$ ($x^2+y^2+z^2<R^2$, where ${\bf x}=(x,y,z)$ is the coordinate and $R$ is the radius) and empty in $x^2+y^2+z^2>R^2$. To get the longest life-time effective theory, expanding $X_{3D}$ at ${\bf k}=(0,0,0)$, we get
     \begin{equation}
   \begin{split}
  iX_{3D}^{eff}({\bf k})=k_x\sigma_x+k_y\sigma_y+k_z\sigma_z.
  \end{split}  \label{X3Deff}
    \end{equation}
It then follows
\begin{eqnarray*}
     & & v_{eff} \\
     &=& \left(\frac{\partial Re(E_{\alpha}^{eff}({\bf k}))}{\partial k_x},\frac{\partial Re(E_{\alpha}^{eff}({\bf k}))}{\partial k_y},\frac{\partial Re(E_{\alpha}^{eff}({\bf k}))}{\partial k_z} \right) \\
     &=& \pm\left(\frac{k_x}{\sqrt{k_x^2+k_y^2+k_z^2}},\frac{k_y}{\sqrt{k_x^2+k_y^2+k_z^2}},\frac{k_z}{\sqrt{k_x^2+k_y^2+k_z^2}}\right),
\end{eqnarray*}
      $|v_{eff}|=1$ and $E_{\alpha}^{eff}$ is the eigenvalues of $iX_{3D}^{eff}({\bf k})$.

   Substituting (\ref{X3Deff}) into Eq.(\ref{GF}) and focusing on the points in the $D^3$,  we get that the wave front after time $t_0$ is a sphere $S^2$ with radius $|R-t_0|$ and center at $(0,0,0)$. The wave front has a Dirac cone structure in the $(3+1)$-dimensional space-time $({\bf x},t)$ (damping wave front equation: $x^2+y^2+z^2=(R-t)^2$ and $t\ge R$). We dub this damping behavior as a 3D Dirac damping since the damping wave front has a Dirac cone structure.

\section{Prove the equivalence of Eq.[\ref{ipm1}] and Eq.[\ref{ipm2}]}

In this appendix, we prove the equivalence of Eq.[\ref{ipm1}] and Eq.[\ref{ipm2}]. In Schr\"{o}dinger picture, the operators do not evolve with time and the density matrix satisfies Eq.(\ref{lindbladeq}). The solution of Lindblad equation can be formally represented as
\[
\rho(t)=e^{\mathcal{L}t}[\rho],
\]
where
\[
e^{\mathcal{L}t}=\sum_{n=0}^{\infty}\frac{(\mathcal{L}t)^n}{n!}.
\]
In the Heisenberg picture, the density matrix does not evolve with time and the operators $O$ satisfy Eq.(\ref{operatoreq}). It follows
\[
O(t)=e^{\mathcal{L}^{\dagger}t}[O],
\]
where
\[
e^{\mathcal{L}^{\dagger}t}=\sum_{n=0}^{\infty}\frac{(\mathcal{L}^{\dagger}t)^n}{n!}.
\]
In the main text,
\[
\rho_{NESS}=|NESS\rangle\langle NESS|=|0\rangle\langle0|
\]
is the density matrix without any particle.

From Eq.(\ref{ipm2}), we have
    \begin{equation}
   \begin{split}
    I_{+}=&|\langle0| (c_{j_2}(t)c^{\dagger}_{j_1}(0)+c^{\dagger}_{j_1}(0)c_{j_2}(t))|0\rangle|^2 \\
           =&|\langle 0| c_{j_2}(t)c^{\dagger}_{j_1}(0)|0\rangle|^2  \\
           =&|\langle0| e^{\mathcal{L}^{\dagger}t}[c_{j_2}(0)]|j_1\rangle|^2  \\
           =&|Tr\left\{|j_1\rangle \langle 0| e^{\mathcal{L}^{\dagger}t}[c_{j_2}(0)]\right\}|^2  \\
           =&|Tr\left\{e^{\mathcal{L}t}[|j_1\rangle \langle0|] c_{j_2}(0)\right\}|^2  \\
           =&|Tr\left\{e^{-iH_{eff}t}|j_1\rangle \langle0| c_{j_2}(0)\right\}|^2  \\
           =&|Tr\left\{e^{-iH_{eff}t}|j_1\rangle \langle j_2| \right\}|^2  \\
           =&|\langle j_2|e^{-iH_{eff}t}|j_1\rangle |^2 .
   \end{split} \label{ip2}
   \end{equation}
    Similarly,
   \begin{equation}
   \begin{split}
    I_{-}=|\langle j_1|e^{-iH_{eff}t}|j_2\rangle |^2 ,
   \end{split}   \label{im2}
   \end{equation}
 where $H_{eff}=H-i\sum_{x=1}^{L}L^{l\dagger}_xL^l_x$, $L$ is the system size, and $L^l_x$ is the Lindblad operator. In the single particle basis $|1A\rangle,|1B\rangle ,|2A\rangle,|2B\rangle,...,|LA\rangle,|LB\rangle $, Eqs.(\ref{ip2}) and (\ref{im2}) are equivalent to Eq.(\ref{ipm1}) (By expanding Eq.(\ref{ipm1}) in real space and taking complex conjugate on the two side of Eq.(\ref{ipm1}), it can be verified.). In the derivation of Eq.(\ref{ip2}), we have used two relations:
  \begin{equation}
  Tr[\hat{P}e^{\mathcal{L}^{\dagger}t}[\hat{Q}]]=Tr[e^{\mathcal{L}t}[\hat{P}]\hat{Q}], \label{s2}
  \end{equation}
  and
   \begin{equation}
  e^{\mathcal{L}t}[|j_1\rangle \langle0|]=e^{-iH_{eff}t}|j_1\rangle \langle 0| . \label{s3}
  \end{equation}

  Proof of Eq.(\ref{s2}): It is easy to verify that $ Tr[\hat{P}\mathcal{L}^{\dagger}[\hat{Q}]]=Tr[\mathcal{L}[\hat{P}]\hat{Q}]$, then we have $Tr[\hat{P}\mathcal{L}^{\dagger n}[\hat{Q}]]=Tr[\mathcal{L}^n[\hat{P}]\hat{Q}]$. It follows
  \begin{equation}
  \begin{split}
  Tr[\hat{P}e^{\mathcal{L}^{\dagger}t}[\hat{Q}]]=&\sum_{n=0}^{\infty}\frac{t^n}{n!} Tr[\hat{P}\mathcal{L}^{\dagger n}[\hat{Q}]]  \\
                                    =&\sum_{n=0}^{\infty}\frac{t^n}{n!} Tr[\mathcal{L}^{n}[\hat{P}]\hat{Q}]  \\
                                    =&Tr[e^{\mathcal{L}t}[\hat{P}]\hat{Q}] .
  \end{split} \label{s4}
  \end{equation}

  Proof of Eq.(\ref{s3}): We begin with
  \begin{equation}
  \begin{split}
  \mathcal{L}[|j_1\rangle \langle 0|]=&-i[H,|j_1\rangle \langle 0|]+\sum_x( 2L^l_x|j_1\rangle \langle 0|L^{l\dagger}_x\\
  &-\left\{L^{l\dagger}_xL^l_x, |j_1\rangle \langle 0| \right\})   \\
                       =&(-iH-\sum_x L^{l\dagger}_xL^l_x)|j_1\rangle \langle 0| .
  \end{split}  \label{s5}
  \end{equation}
     Here we have used $\langle 0|H=0$ and $\langle 0|L^{l\dagger}_x=0$. Assume that
       \begin{equation}
  \begin{split}
  \mathcal{L}^n[|j_1\rangle \langle 0|]=(-iH-\sum_x L^{l\dagger}_xL^l_x)^n|j_1\rangle \langle 0|
  \end{split} , \label{s6}
  \end{equation}
  then we get
    \begin{equation}
  \begin{split}
  \mathcal{L}^{n+1}[|j_1\rangle \langle 0|]=&\mathcal{L}[(-iH-\sum_x L^{l\dagger}_xL^l_x)^n|j_1\rangle\langle0|] \\
                             =&-i[H,(-iH-\sum_x L^{l\dagger}_xL^l_x)^n|j_1\rangle \langle0|] \\
                             &+\sum_x\left( 2L^l_x(-iH-\sum_x L^{l\dagger}_xL^l_x)^n|j_1\rangle \langle 0|L^{l\dagger}_x    \right. \\
                              -&\left. \left\{L^{l\dagger}_xL^l_x, (-iH-\sum_x L^{l\dagger}_xL^l_x)^n|j_1\rangle \langle0| \right\}\right)   \\
                             =&(-iH-\sum_x L^{l\dagger}_xL^l_x)^{n+1}|j_1\rangle \langle 0| .
  \end{split}  \label{s7}
  \end{equation}
   Combining Eqs.(\ref{s5}), (\ref{s6}) and (\ref{s7}), we conclude that Eq.(\ref{s6}) holds true for any $n$. Finally we get
   \begin{equation}
   \begin{split}
   e^{\mathcal{L}t}[|j_1\rangle\langle0|]=&\sum_{n=0}^{\infty}\frac{t^n}{n!}\mathcal{L}^n[|j_1\rangle\langle0|]  \\
                             =&\sum_{n=0}^{\infty}\frac{t^n}{n!}(-iH-\sum_x L^{l\dagger}_xL^l_x)^n|j_1\rangle\langle0| \\
                             =&e^{-iH_{eff}t}|j_1\rangle\langle0| .
   \end{split}
   \end{equation}

\section{The analytical representation of $I_C$ for general models}

 In this section, we derive the analytical representation of
 \begin{equation}
   \begin{split}
   I_C(j_1,j_2,t)=&\frac{I_+}{I_-}
   \end{split} \label{Ir}
 \end{equation}
 for a general 1D quadratic Lindbladian system. And we also give the analytical representation of Eq. (\ref{Ir}) for a general d-dimensional quadratic Lindbladian system.

 The Green function $\Delta_{ij}=Tr(\rho c_i^{\dagger}c_j)$ of the system is governed by
   \[
   \tilde{\Delta}=\Delta-\Delta_s=e^{Xt}\tilde{\Delta}(0)e^{X^{\dagger}t},
   \]
   where $\Delta_s$ is the steady value of $\Delta$ and $X$ is the damping matrix of a general 1D model with the matrix in the momentum space given by $X(k)$. In our main text, $X$ is effectively described by a non-Hermitian SSH model. To get the analytical representation of Eq. (\ref{Ir}), we should derive the analytical representation of $\langle \langle j_2 |e^{Xt}|j_1 \rangle \rangle $, where $j_1,j_2={11,12,...,21,22,...,PQ}$, $P$ is the number of cells and $Q$ is the total inner degrees of freedom in the cell. For convenience, we denote
   \begin{equation}
\begin{split}
T(j_1,j_2,t)=\langle \langle j_2 |e^{Xt}|j_1 \rangle \rangle
\end{split} \label{T}
\end{equation}
with $j_1=x_1q_1$, $j_2=x_2q_2$, where $x_1,x_2={1,2,...,P}$ is the cell index and $q_1,q_2={1,2,...,Q}$ is the index of the inner degree of freedom in the cell. We will derive its analytical representation of $T(j_1,j_2,t)$ and Eq. (\ref{Ir}) under both PBC and OBC.

   \subsection{PBC case}
    Assume that $E_{\alpha}(k)$, $|k,\alpha\rangle \rangle_R$ and $|k,\alpha\rangle \rangle_L$ are the eigenvalues, right eigenvectors and left eigenvectors of $X(k)$, where $\alpha$ is the band index and $k$ is the momentum, $|k ,\alpha\rangle \rangle_R=|k\rangle \rangle \otimes |\psi(k,\alpha)\rangle \rangle_R$, $_ L\langle \langle k ,\alpha|=\langle \langle k| \otimes$$_L\langle \langle\psi(k,\alpha)|$,$|x_1q_1\rangle \rangle=|x_1\rangle \rangle \otimes |q_1\rangle \rangle$ and $|x_2q_2\rangle \rangle=|x_2\rangle \rangle \otimes |q_2\rangle \rangle$. While $|\psi(k,\alpha)\rangle \rangle_R$, $|\psi(k,\alpha)\rangle \rangle_L$, $|q_2\rangle \rangle$ and $|q_2\rangle \rangle$ belong to the Hilbert space in the unit cell, $|k\rangle \rangle$, $|x_1\rangle \rangle$ and $|x_2\rangle \rangle$ belong to the Hilbert space of cell index. We have $\langle \langle x_1|k\rangle \rangle=e^{ik}$, $\langle \langle x_1q_1|k,\alpha\rangle \rangle_R=\langle \langle x_1| k \rangle \rangle \langle \langle q_1| \psi(k,\alpha) \rangle \rangle_R=e^{ikx_1}\langle \langle q_1| \psi(k,\alpha) \rangle \rangle_R$ and $\sum_{k',\alpha'}|k',\alpha'\rangle \rangle_{RL}\langle \langle k',\alpha'|=\mathbb{I}$. It follows
   \begin{equation}
\begin{split}
&T(j_1,j_2,t)\\
=&\sum_k \langle \langle j_2 |e^{X(k)t}|j_1 \rangle \rangle  \\
=&\sum_{k,k',\alpha'} \langle \langle j_2 |e^{X(k)t}|k',\alpha'\rangle \rangle_{RL}\langle \langle k',\alpha'|j_1 \rangle \rangle        \\
=&\sum_{k,\alpha} \langle \langle j_2 |k,\alpha\rangle \rangle_{RL}\langle \langle k,\alpha|j_1 \rangle \rangle e^{E_{\alpha}(k)t} \\
=&\sum_{k,\alpha} \langle \langle q_2 |\psi(k,\alpha)\rangle \rangle_{RL}\langle \langle \psi(k,\alpha)|q_1 \rangle \rangle e^{E_{\alpha}(k)t+ikx_2-ikx_1} .  \label{Tpbc1}
\end{split}
\end{equation}
 Substituting $x_2-x_1=v_{\alpha}(k)t$ with $v_{\alpha}(k)=\frac{\partial Re[iE_{\alpha}(k)]}{\partial k}$ into the above expression, we get
   \begin{equation}
\begin{split}
&T(j_1,j_2,t)\\
=&\sum_{k,\alpha} \langle \langle q_2 |\psi(k,\alpha)\rangle \rangle_{RL}\langle \langle \psi(k,\alpha)|q_1 \rangle \rangle e^{E_{\alpha}(k)t+ikv_{\alpha}(k)t}. \label{Tpbc2}
\end{split}
\end{equation}
Substituting Eqs.(\ref{T}) and (\ref{ipm1}) into Eq.(\ref{Ir}), we get the analytical representation of Eq.(\ref{Ir}):
\begin{equation}
\begin{split}
  &I_C(j_1,j_2,t) \\
  =&T(j_1,j_2,t)T^{\dagger}(j_1,j_2,t)/(T(j_2,j_1,t)T^{\dagger}(j_2,j_1,t))  \\
       =&|T(j_1,j_2,t)|^2/|T(j_2,j_1,t)|^2 . \label{Irt}
\end{split}
\end{equation}

 {\bf General $d$-dimensional model.} Similarly, for a general $d$-dimensional model, we can get the analytical representation of Eq.(\ref{Ir}),
 \begin{equation}
\begin{split}
  I_C(j_1,j_2,t)=&|T(j_1,j_2,t)|^2/|T(j_2,j_1,t)|^2
\end{split} \label{Ird}
\end{equation}
with
\begin{equation}
\begin{split}
&T(j_1,j_2,t)\\
=&\sum_{{\bf k}}\langle \langle j_2 |e^{X({\bf k})t}|j_1 \rangle \rangle  \\
=&\sum_{{\bf k},\alpha} \langle \langle q_2 |\psi({\bf k},\alpha)\rangle \rangle_{RL}\langle \langle \psi({\bf k},\alpha)|q_1 \rangle \rangle e^{E_{\alpha}({\bf k})t+i{\bf k}({\bf x}_2-{\bf x}_1)},
\end{split} \label{Tpbcd}
\end{equation}
where $j_1={\bf x}_1q_1,j_2={\bf x}_2q_2$, ${\bf x}_1$ and ${\bf x}_2$ are $d$-dimensional vectors which label the location of cells, $q_1$ and $q_2$ label the degree of freedom in the cell, ${\bf k}$ is a $d$-dimensional momentum, and $\alpha$ is the band index of $X({\bf k})$. $E_{\alpha}({\bf k})$, $|{\bf k},\alpha\rangle \rangle_R$ and $|{\bf k},\alpha\rangle \rangle_L$ are the eigenvalues, right eigenvectors and left eigenvectors of $X({\bf k})$, respectively. $|{\bf k}, \alpha\rangle \rangle_R=|{\bf k}\rangle \rangle \otimes |\psi({\bf k}, \alpha)\rangle \rangle_R$,  $_ L\langle \langle {\bf k}$, $\alpha|=\langle \langle {\bf k}| \otimes$$_L\langle \langle\psi({\bf k}$, $\alpha)|$, $|x_1q_1\rangle \rangle=|{\bf x}_1\rangle \rangle \otimes |q_1\rangle \rangle$ and $|{\bf x}_2q_2\rangle \rangle=|{\bf x}_2\rangle \rangle \otimes |q_2\rangle \rangle$. While $|\psi({\bf k},\alpha)\rangle \rangle_R$, $|\psi({\bf k},\alpha)\rangle \rangle_L$, $|q_2\rangle \rangle$ and $|q_2\rangle \rangle$ belong to the Hilbert space in the unit cell, $|k\rangle \rangle$, $|{\bf x}_1\rangle \rangle$ and $|{\bf x}_2\rangle \rangle$ belong to the Hilbert space of cell index.

 {\bf Example:} Here, we apply this formula to the model discussed in the main text. For this model, the damping matrix is given by
 \begin{equation}
\begin{split}
  X(k)=i[t_1+t_2 \cos(k)]\sigma_x+it_2 \sin (k)\sigma_y+\frac{\gamma}{2}\sigma_y-\frac{\gamma}{2}\sigma_0 , \label{sshk}
\end{split}
\end{equation}
and we have
 \begin{equation}
\begin{split}
E_{\pm}(k)=-\frac{\gamma}{2}\pm i\sqrt{t_1^2+t_2^2+2t_1t_2 \cos (k)-\frac{\gamma^2}{4}-i\gamma t_2 \sin (k)}
\end{split}  \label{sshe}
\end{equation}
and $v_{\pm}(k)=\partial(Re[iE_{\pm}(k)])/\partial k$. For the parameter set the same as in the main text, it can be verified that $max(Re(E))\approx Re(E_{-}(\pi))=0$, $max(v)\approx v_{-}(\pi)=1$, $min(Re(E))\approx Re(E_{+}(\pi))=-0.8$, and $min(v)\approx v_{+}(\pi)=-1$.

 Substituting $t=j_2-j_1=m$ and $j_1=25B$ into Eq.(\ref{Tpbc2}), we get
 \begin{equation}
\begin{split}
&|T(j_1,j_2,t)||_{t=j_2-j_1}\\
\approx & \langle \langle B |\psi(\pi,-)\rangle \rangle_{RL}\langle \langle \psi(\pi,-)|B \rangle \rangle e^{E_{-}(\pi)t+i\pi v_{-}(\pi) t} .
\end{split}  \label{Tpssh1}
\end{equation}
Here we have used that $x_2-x_1=tv_{\alpha}(k)$, thus $v_{\alpha}(k)=(x_2-x_1)/t=1$.  We can get that $\alpha=-$ and $k\approx \pi$.
Similarly, for $|T(j_2,j_1,t)|$, $t=j_2-j_1=m$ and $j_1=25B$, we have
\begin{equation}
\begin{split}
&|T(j_2,j_1,t)||_{t=j_2-j_1}\\
\approx & \langle \langle B |\psi(\pi,+)\rangle \rangle_{RL}\langle \langle \psi(\pi,+)|B \rangle \rangle e^{E_{+}(\pi)t+i\pi v_{+}(\pi) t} .
\end{split} \label{Tpssh2}
\end{equation}
Here we have used that $x_1-x_2=tv_{\alpha}(k)$, thus $v_{\alpha}(k)=(x_1-x_2)/t=-1$. We can get that $\alpha=+$ and $k\approx \pi$.
Substituting Eqs.(\ref{Tpssh1}) and (\ref{Tpssh2}) into Eq.(\ref{Irt}),
we get
\begin{equation}
\begin{split}
&I_C(j_1,j_2,t)|_{t=j_2-j_1}\\
\approx &\left| \frac{\langle \langle B |\psi(\pi,-)\rangle \rangle_{RL}\langle \langle \psi(\pi,-)|B \rangle \rangle}
{\langle \langle B |\psi(\pi,+)\rangle \rangle_{RL}\langle \langle \psi(\pi,+)|B \rangle \rangle} \right|^2 e^{1.6t} \\
\approx & e^{1.6t} \\
=& e^{1.6m} .
\end{split} \label{Irpssh}
\end{equation}
It is consistent with the result in the main text (For this special model we get $I_C$ for the system under OBC in the main text).
 \subsection{OBC case}
  In this subsection, the non-Bloch band theory is applied to get the analytical representation of Eq.(\ref{Ir}). Assume that the GBZ of $X$ is $z=e^{i(k+i \kappa)}$, where $z$ and $\kappa$ is a function of $k$ and the band index $\alpha$ (see Ref. \cite{KYokomizo,YYi} for methods to obtain the GBZ of 1D systems). For convenience, we use $\kappa$ representing $\kappa(k,\alpha)$, and $E(k+i\kappa)$, $|k+i\kappa,\alpha \rangle \rangle_R$ and $|k+i\kappa,\alpha\rangle \rangle_L$ denoting the eigenvalues, right eigenvectors and left eigenvectors of $X(k+i\kappa)$, respectively. Here, $|k+i\kappa ,\alpha\rangle \rangle_R=|k+i\kappa\rangle \rangle_R \otimes |\psi(k+i\kappa,\alpha)\rangle \rangle_R$, $_ L\langle \langle k+i\kappa,\alpha|=$$_L\langle \langle k+i\kappa| \otimes$$_L\langle \langle\psi(k+i\kappa,\alpha)|$,$|x_1q_1\rangle \rangle=|x_1\rangle \rangle \otimes |q_1\rangle \rangle$ and $|x_2q_2\rangle \rangle=|x_2\rangle \rangle \otimes |q_2\rangle \rangle$.
  While
  $|\psi(k+i\kappa,\alpha)\rangle \rangle_R$, $|\psi(k+i\kappa,\alpha)\rangle \rangle_L$, $|q_2\rangle \rangle$ and $|q_2\rangle \rangle$ belong to the Hilbert space in the unit cell, $|k+i\kappa\rangle \rangle_R$, $|k+i\kappa\rangle \rangle_L$, $|x_1\rangle \rangle$ and $|x_2\rangle \rangle$ belong to the Hilbert space of cell index. We have $\langle \langle x_1|k+i\kappa\rangle \rangle_R=e^{i(k+i\kappa)}$, $\langle \langle x_1|k+i\kappa\rangle \rangle_L=e^{i(k-i\kappa)}$, $\langle \langle x_1q_1|k+i\kappa,\alpha\rangle \rangle_R=\langle \langle x_1| k+i\kappa \rangle \rangle_R \langle \langle q_1| \psi(k+i\kappa,\alpha) \rangle \rangle_R=e^{i(k+i\kappa)x_1}\langle \langle q_1| \psi(k+i\kappa,\alpha) \rangle \rangle_R$, $\sum_{k',\alpha'}|k'+i\kappa',\alpha'\rangle \rangle_{RL}\langle \langle k'+i\kappa',\alpha'|=\mathbb{I}$. Under OBC, $\langle \langle x_2q_2 |e^{Xt}|x_1q_1 \rangle \rangle$ can be decomposed to each GBZ modes $\langle \langle x_2q_2 |e^{X(k+i\kappa)t}|x_1q_1 \rangle \rangle$, i.e., $\langle \langle x_2q_2 |e^{Xt}|x_1q_1 \rangle \rangle=\sum_{k}\langle \langle x_2q_2 |e^{X(k+i\kappa)t}|x_1q_1 \rangle \rangle$. Taking account into these and substituting $j_1=x_1q_1$, $j_2=x_2q_2$ into Eq.(\ref{T}), we get
     \begin{equation}
\begin{split}
&T(j_1,j_2,t)\\
=&\sum_{k} \langle \langle x_2q_2 |e^{X(k+i\kappa)t}|x_1q_1 \rangle \rangle  \\
       =&\sum_{k,k',\alpha'} \langle \langle x_2q_2 |e^{X(k+i\kappa)t}|k'+i\kappa',\alpha'\rangle \rangle_{R}\times\\
       &_{L}\langle \langle k'+i\kappa',\alpha'|x_1q_1 \rangle \rangle  \\
       =&\sum_{k,\alpha} \langle \langle x_2q_2 |k+i\kappa,\alpha \rangle \rangle_{RL}\langle \langle k+i\kappa,\alpha|x_1q_1 \rangle \rangle e^{E(k+i\kappa)t} \\
       =&\sum_{k,\alpha} \langle \langle q_2 |\psi(k+i\kappa,\alpha) \rangle \rangle_{R}\times\\
       &_{L}\langle \langle \psi(k+i\kappa,\alpha) |q_1\rangle \rangle e^{E(k+i\kappa)t+ix_2(k+i\kappa)-ix_1(k+i\kappa)}  \\
       =&\sum_{k,\alpha} \langle \langle q_2 |\psi(k+i\kappa,\alpha) \rangle \rangle_{R}\times\\
       &_{L}\langle \langle \psi(k+i\kappa,\alpha) |q_1\rangle \rangle e^{E(k+i\kappa)t+i(x_2-x_1)(k+i\kappa)} .
\end{split} \label{Tobc}
\end{equation}
 Substituting it into Eq.(\ref{Irt}), we get the analytical representation of Eq.(\ref{Ir}).

  {\bf Example:} Here, we apply this formula to the model discussed in the main text, with the damping matrix given by Eq.(\ref{sshk}). We have $v_{\pm}(k+i\kappa)=\partial(Re[iE_{\pm}(k+i\kappa)])/\partial (k+i\kappa)$. For the parameter set the same as the main text, it can be verified that the GBZ of this model is $z=e^{i(k-0.42i)}$ ($k\in [0,2\pi)$), and we have $Re(E_{\pm}(k-0.42i))=-0.4$, $max(v)\approx v_{-}(\pi-0.42i)=1$ and  $min(v)\approx v_{+}(\pi-0.42i)=-1$.

   Substituting $t=j_2-j_1=x_2-x_1=m$ and $j_1=25B$ into Eq.(\ref{Tobc}), we get
      \begin{equation}
\begin{split}
&T(j_1,j_2,t)|_{t=j_2-j_1}\\
\approx & \langle \langle B |\psi(\pi-0.42i,-) \rangle \rangle_{R}\times\\
       &_{L}\langle \langle \psi(\pi-0.42i,-) |B\rangle \rangle e^{E_{-}(\pi-0.42i)t+im(\pi-i0.42)} .
\end{split} \label{Tossh1}
\end{equation}
Here we use that $x_2-x_1=tv_{\alpha}(k-0.42i)$, thus $v_{\alpha}(k-0.42i)=(x_2-x_1)/t=1$. We can get that $\alpha=-$ and $k\approx \pi$.
 Similarly, for $T(j_2,j_1,t)$, $t=j_2-j_1=x_2-x_1=m$ and $j_1=25B$, we have
 \begin{equation}
\begin{split}
&T(j_2,j_1,t)|_{t=j_2-j_1} \\
\approx & \langle \langle B |\psi(\pi-0.42i,+) \rangle \rangle_{R}\times\\
       &_{L}\langle \langle \psi(\pi-0.42i,+) |B\rangle \rangle e^{E_{+}(\pi-0.42i)t-im(\pi-i0.42)} .
\end{split} \label{Tossh2}
\end{equation}
Here we use that $x_1-x_2=tv_{\alpha}(k-0.42i)$, thus $v_{\alpha}(k-0.42i)=(x_1-x_2)/t=-1$. We can get that $\alpha=+$ and $k\approx \pi$.

Substituting Eqs.(\ref{Tossh1})(\ref{Tossh2}) and $Re(E_{\pm}(k-0.42i))=-0.4$ into Eq.(\ref{Irt}), we have
 \begin{equation}
\begin{split}
&I_C(j_1,j_2,t)|_{t=j_2-j_1} \\
\approx &\left| \frac{\langle \langle B |\psi(\pi-0.42i,-) \rangle \rangle_{RL}\langle \langle \psi(\pi-0.42i,-) |B\rangle \rangle }
{\langle \langle B |\psi(\pi-0.42i,+) \rangle \rangle_{RL}\langle \langle \psi(\pi-0.42i,+) |B\rangle \rangle } \right|^2e^{1.68m}  \\
\approx& e^{1.68m} .
\end{split}  \label{Irossh}
\end{equation}
It is consistent with the result in the main text.

\section{Analytical derivation of chiral damping and helical damping via information constraint}
In this section, we give the analytical derivation of chiral damping and helical damping via information constraint.

\subsection{Chiral damping}
We have given the analytical representation of $I_C$ for the general $d$-dimensional quadratic Lindbladian system in the previous appendix. For the model in the main text with parameters set as $t_1=t_2=1$ and $\gamma=0.8$, we get that $I_C(j_1,j_2,t)$ takes the following form
\begin{equation}
I_C(j_1,j_2,t)=\frac{|G_{j_1j_2}(t)|^2}{|G_{j_2j_1(t)}|^2}\approx e^{1.6(x_2-x_1)} , \label{Irmt}
\end{equation}
where $j_1=x_1q_1$, $j_2=x_2q_2$, $x_1$ and $x_2$ are the cell indexes with $1\le x_1\le x_2\le L$
, $q_1,q_2 \in \left\{ A,B\right\}$ label the degree of freedom in the cell, and $L$ is the size of system.
 Considering the case with fully filled initial state, here we derive the analytical representation of the Green function. From previous section, we know that $|G_{j_1j_2}(t)|^2=|T(j_1,j_2,t)|^2$. According to Eq.(\ref{Tpbc2}),  $|G_{j_1j_2}(t)|^2$ is dominated by the term with largest real part of $E_{\alpha}(k)$. Here $E_{\alpha}(k)$ is the eigenvalues of $iX(k)$, where $\alpha$ is the band index, $k$ is the momentum and $i$ is the imaginary unit. Eq.(\ref{sshe}) is the expression of $E_{\alpha}(k)$. Noticing that $max(Re(E_{\alpha}(k))=Re(E_{-}(\pi))=0$, we have
\begin{equation}
\begin{split}
&|G_{j_1j_2}(t)|^2\\
=&|T(j_1,j_2,t)|^2 \\
=&|\sum_{k,\alpha} \langle \langle q_2 |\psi(k,\alpha)\rangle \rangle_{RL}\langle \langle \psi(k,\alpha)|q_1 \rangle \rangle e^{E_{\alpha}(k)t+ikv_{\alpha}(k)t}|^2 \\
\approx &|\langle \langle q_2 |\psi(k,\alpha)\rangle \rangle_{R}\times\\
       &_{L}\langle \langle \psi(k,\alpha)|q_1 \rangle \rangle |_{\alpha=-,k=\pi} e^{E_{-}(\pi)t+ikv_{-}(\pi)t}|^2, \label{Gj1j2a}
\end{split}
\end{equation}
where $v_{\alpha}(k)=\partial(Re[iE_{\alpha}(k)])/\partial k$, and $v_{\alpha}(k)$ also satisfies the constraint $x_2-x_1-v_{\alpha}(k)t=0$. Substituting $v_{-}(\pi)=1$ and $x_2-x_1-v_{\alpha}(k)t=0$ into Eq.({\ref{Gj1j2a}}), we get that
\begin{equation}
|G_{j_1j_2}(t)|^2\approx f_{q_1q_2} \delta(x_2-x_1-t) , \label{Gj1j2b}
\end{equation}
where $f_{q_1q_2}=\langle \langle q_2 |\psi(k,\alpha)\rangle \rangle_{RL}\langle \langle \psi(k,\alpha)|q_1 \rangle \rangle |_{\alpha=-,k=\pi}$, $\delta(0)=1$ and $\delta(x)=0$ when $x\ne 0$. Substituting Eq.(\ref{Gj1j2b}) into Eq.(\ref{Irmt}), we get
\begin{equation}
|G_{j_2j_1}(t)|^2=f_{q_1q_2} \delta(x_2-x_1-t)e^{-1.6(x_2-x_1)} . \label{Gj2j1a}
\end{equation}
  According to the definition of Green function,  $|G_{j_1j_2}(t)|^2$ is the probability for creating a particle at the space-time $(j_1,0)$ and annihilating at $(j_2,t)$ \cite{PeskinAndSchroeder}. Thus, we get $n_{xA} (t)$ under the OBC ($n_{xA} (t)$ is the total particle number at $x$ cell and $A$ site):
\begin{equation}
\begin{split}
&n_{xA}(t)\approx \sum_{x_3=1}^L\sum_{q_3=A,B}|G_{x_3q_3,xA}|^2\\
&+\sum_{x_3=1}^L\sum_{q_3=A,B}
\int_0^t dt_1 |G_{x_3q_3,1A}(t_1)|^2|G_{1A,xA}(t-t_1)|^2\\
&+\sum_{x_3=1}^L\sum_{q_3=A,B}
\int_0^t dt_1 |G_{x_3q_3,LB}(t_1)|^2|G_{LB,xA}(t-t_1)|^2,
\end{split} \label{nxA}
\end{equation}
where the first term is the contribution of reflectionless wave (zero order), and the second and third terms are the contributions of primary scattering wave (first order) at left and right boundary, respectively. We consider the case $t<\frac{L}{v}$  ($v=max(v_{\alpha}(k))=1$ is the maximum velocity), and thus there is no contribution of high-order scattering waves. Substituting Eq.(\ref{Gj1j2b}) and Eq.(\ref{Gj2j1a}) into Eq.(\ref{nxA}), we get
\begin{equation}
\begin{split}
n_{xA}(t)\approx &\Theta_1(x-1-t)(f_{AA}+f_{BA})\\
&+\Theta_1(L-x-t)(f_{AB}+f_{AA})e^{-1.6t}\\
 &+\Theta_2(t-x+1)(f_{AB}+f_{AA})f_{AA}
 e^{-1.6(t-x)}\\
 &+\Theta_2(t+x-L)(f_{AB}+f_{BB})f_{AB}
 e^{-1.6(L-x)},
\end{split}
\end{equation}
where $\Theta_1(y)$ and $\Theta_2(y)$ are Heaviside step functions, $\Theta_1(y)=1$ for $y\ge 0$ and $\Theta_1(y)=0$ for $y<0$, $\Theta_2(y)=1$ for $y>0$ and $\Theta_2(y)=0$ for $y\le 0$. Thus, the damping wave-front equation is $x-1-t =0$. The damping wave-front is of a 1D chiral Dirac fermion structure.

\subsection{Helical damping}
We consider the model discussed in Ref.\cite{CHLiu3} with parameters set as $t_1=t_2=1$, $\gamma_l=0.8$, $\gamma_g=0$ and $\delta_1=\delta_2=0$, and the system is initially fully filled. This model is a combination of two decoupled models in the main text with opposite propagating directions. Using the above conclusions for chiral damping, we get
\begin{equation}
I_C(j_1\uparrow,j_2\uparrow,t)=\frac{|G_{j_1\uparrow j_2\uparrow }(t)|^2}{|G_{j_2\uparrow j_1\uparrow (t)}|^2}\approx e^{1.6(x_2-x_1)},
\end{equation}
\begin{equation}
|G_{j_1\uparrow j_2\uparrow}(t)|^2\approx f_{q_1q_2} \delta(x_2-x_1-t),
\end{equation}
\begin{equation}
|G_{j_2\uparrow j_1\uparrow}(t)|^2=f_{q_1q_2} \delta(x_2-x_1-t)e^{-1.6(x_2-x_1)} ,
\end{equation}
\begin{equation}
I_C(j_1\downarrow,j_2\downarrow,t)=\frac{|G_{j_1\downarrow j_2\downarrow}(t)|^2}{|G_{j_2\downarrow j_1\downarrow (t)}|^2}\approx e^{-1.6(x_2-x_1)},
\end{equation}
\begin{equation}
|G_{j_1\downarrow j_2\downarrow}(t)|^2\approx f_{q_1q_2} \delta(x_2-x_1-t)e^{-1.6(x_2-x_1)},
\end{equation}
\begin{equation}
|G_{j_2\downarrow j_1\downarrow}(t)|^2=f_{q_1q_2} \delta(x_2-x_1-t),
\end{equation}
and
\begin{equation}
\begin{split}
n_{xA}(t) &=n_{xA\uparrow}(t)+n_{xA\downarrow}(t)  \\
=&n_{xA\uparrow}(t)+n_{(L-x)A\uparrow}(t)  \\
 \approx &\Theta_1(x-1-t)(f_{AA}+f_{BA})\\
 &+\Theta_1(L-x-t)(f_{AB}+f_{AA})e^{-1.6t}\\
 &+\Theta_2(t-x+1)(f_{AB}+f_{AA})f_{AA}e^{-1.6(t-x)}\\
 &+\Theta_2(t+x-L)(f_{AB}+f_{BB})f_{AB}e^{-1.6(L-x)} \\
 &+\Theta_1(L-x-1-t)(f_{AA}+f_{BA})\\
 &+\Theta_1(x-t)(f_{AB}+f_{AA})e^{-1.6t}\\
 &+\Theta_2(t-L+x+1)(f_{AB}+f_{AA})f_{AA}
 e^{-1.6(t-L+x)}\\
 &+\Theta_2(t-x)(f_{AB}+f_{BB})f_{AB}
 e^{-1.6x},
\end{split}
\end{equation}
where $n_{xA\uparrow}(t)$ ($n_{xA\downarrow}(t)$) is the total spin up (down) particle number at the $x$ cell and $A$ site.
The wave-front equations are $x-1-t =0$ and $L-x-1-t =0$. The damping wave-front is of a 1D helical Dirac fermion structure.

\section{Proof of ``If $X^{T}=X$, then $I_C=1$"}
{\bf Proposition:} For a quadratic Lindbladian system, if the damping matrix $X$ satisfies $X^{T}=X$, then $I_C=1$.

Here we give the proof of this proposition. For a general $d$-dimensional quadratic Lindbladian system, assume that $j_1={\bf x}_1q_1,j_2={\bf x}_2q_2$, ${\bf x}_1$ and ${\bf x}_2$ are $d$-dimensional vectors labeling location of cells, $q_1$ and $q_2$ are the indexes that label the degree of freedom in the cell. $|j_1 \rangle \rangle$ is a matrix representation of $|j_1\rangle$ in the single particle basis $[|11\rangle,|12\rangle,...,|1Q\rangle,|21\rangle,|22\rangle,...,|2Q\rangle,...,|L1\rangle,|L2\rangle,...,|LQ\rangle]$, where $Q$ is the total degree of freedom in the cell and $L$ is the system size.
 Without loss of generality, we can let $|j_1 \rangle \rangle^*=|j_1 \rangle \rangle$ and $|j_2 \rangle \rangle^*=|j_2 \rangle \rangle$. There exist real numbers $\theta_1$ and $\theta_2$ so that $|j_1 \rangle \rangle$ and $|j_2 \rangle \rangle$ can be real matrices after a gauge transformation $|j_1 \rangle \rangle\rightarrow e^{i\theta_1}|j_1 \rangle \rangle$ and $|j_2 \rangle \rangle\rightarrow e^{i\theta_2}|j_2 \rangle \rangle$.

Substitute $X^{T}=X$, $|j_1 \rangle \rangle^*=|j_1 \rangle \rangle$ and $|j_2 \rangle \rangle^*=|j_2 \rangle \rangle$ into Eq.(\ref{T}).
Since $T(j_1,j_2,t)$ is a number,  we can take transpose on both sides of Eq.(\ref{T}), and thus we have
\begin{equation}
\begin{split}
T(j_1,j_2,t)=&[T(j_1,j_2,t)]^T \\
=&[\langle \langle j_2 |e^{Xt}|j_1 \rangle \rangle]^T \\
=&\langle \langle j_1 |e^{X^Tt}|j_2 \rangle \rangle  \\
=&\langle \langle j_1 |e^{Xt}|j_2 \rangle \rangle  \\
=&T(j_2,j_1,t) .
\end{split}  \label{Tdpm}
\end{equation}
Substituting Eq.(\ref{Tdpm}) into Eq.(\ref{Irt}), we have $I_C=1$.

\twocolumngrid


\begin{thebibliography}{99}

\bibitem{GLindblad}G. Lindblad, On the generators of quantum dynamical semigroups, Commun. Math. Phys. {\bf48}, 119 (1976).
      \bibitem{Molmer}J. Dalibard, Y. Castin, and K. Molmer, Wave-Function Approach to Dissipative Processes in Quantum Optics, Phys. Rev. Lett. {\bf 68}, 580 (1992).
                                          \bibitem{Carmichael}H.J. Carmichael, Quantum Trajectory Theory for Cascaded Open Systems, Phys. Rev. Lett. {\bf 70}, 2273 (1993).
                                              \bibitem{Daley}A.J. Daley, Quantum trajectories and open many-body quantum systems, Adv. Phys. {\bf 63}, 77 (2014).


\bibitem{Zoller} S. Diehl, A. Micheli, A. Kantian, B. Kraus, H. P. B\"{u}chler, and P. Zoller, Quantum states and phases in driven open quantum systems with cold atoms, Nat. Phys. {\bf 4}, 878 (2008).

%
\bibitem{TProsen1} T. Prosen, Third quantization: a general method to solve master equations for quadratic open Fermi systems,  New J. Phys. {\bf10}, 043026 (2008).
\bibitem{TProsen2} T. Prosen, Spectral theorem for the Lindblad equation for quadratic open fermionic systems, J. Stat. Mech. (2010) P07020.



\bibitem{TProsen3}T. Prosen, and E. Ilievski, Nonequilibrium phase transition in a periodically driven xy spin chain, Phys. Rev. Lett. {\bf 107}, 060403 (2011).
\bibitem{SLieu}S. Lieu, M. McGinley, and N. R. Cooper, Tenfold Way for Quadratic Lindbladians, Phys. Rev. Lett. {\bf 124}, 040401 (2020).
    \bibitem{HShen}H. Shen, B. Zhen, and L. Fu, Topological Band Theory for Non-Hermitian Hamiltonians, Phys. Rev. Lett. {\bf 120}, 146402 (2018).
    \bibitem{VKozii}V. Kozii and L. Fu, Non-Hermitian topological theory of finite-lifetime quasiparticles: Prediction of bulk Fermi arc due to exceptional point, arXiv:1708.05841.

\bibitem{TELee}T. E. Lee, Anomalous edge state in a non-hermitian lattice, Phys. Rev. Lett. {\bf 116}, 133903 (2016).




\bibitem{Alvarez} V. M. Martinez Alvarez, J. E. Barrios Vargas, and L. E. F. Foa Torres, Non-Hermitian robust edge states in one dimension: anomalous localization and eigenspace condensation at exceptional points, Phys. Rev. B \textbf{97}, 121401(R) (2018).

\bibitem{Xiong} Y. Xiong, Why does bulk boundary correspondence fail in some
non-Hermitian topological models, J. Phys. Commun. {\bf 2}, 035043 (2018).

\bibitem{SYao1} S. Yao and Z. Wang, Edge States and Topological Invariants of Non-Hermitian Systems, Phys. Rev. Lett. {\bf 121}, 086803 (2018).



\bibitem{Kunst} F. K. Kunst, E. Edvardsson, J. C. Budich, and E. J. Bergholtz, Biorthogonal bulk-boundary correspondence in
non-Hermitian systems, Phys. Rev. Lett. \textbf{121}, 026808 (2018).

 \bibitem{SYao2}S. Yao, F. Song, and Z. Wang, Non-Hermitian Chern Bands, Phys. Rev. Lett. {\bf 121}, 136802(2018).



 \bibitem{LeeCH} C. H. Lee and R. Thomale, Anatomy of skin modes and topology in non-Hermitian systems, Phys. Rev. B {\bf 99}, 201103(R) (2019).

\bibitem{JiangH} H. Jiang, L. J. Lang, C. Yang., S. L. Zhu, and
S. Chen, Interplay of non-Hermitian skin effects and Anderson localization
in nonreciprocal quasiperiodic lattices, Phys. Rev. B \textbf{100}, 054301
(2019).

 \bibitem{KYokomizo}K. Yokomizo and S. Murakami, Non-Bloch Band Theory of Non-Hermitian Systems, Phys. Rev. Lett. {\bf 123}, 066404 (2019).
 \bibitem{KZhang} K. Zhang, Z. Yang, and C. Fang, Correspondence between winding numbers and skin modes in non-hermitian systems,
                         Phys. Rev. Lett. {\bf 125}, 126402 (2020). 

\bibitem{NOkuma}N. Okuma, K. Kawabata, K. Shiozaki, and M. Sato, Topological Origin of Non-Hermitian Skin Effects, Phys. Rev. Lett. {\bf 124}, 086801 (2020).
\bibitem{Slager} D. S. Borgnia, A. J. Kruchkov, R.-J. Slager, Non-Hermitian Boundary Modes, Phys. Rev. Lett. \textbf{124}, 056802 (2020).
\bibitem{ZSYang} Z. Yang, K. Zhang, C. Fang, and J. Hu, Non-Hermitian Bulk-Boundary Correspondence and Auxiliary Generalized Brillouin Zone Theory, Phys. Rev. Lett. {\bf 125}, 226402(2020).
 \bibitem{YYi}Y. Yi and Z. Yang, Non-Hermitian skin modes induced by on-site dissipations and chiral tunneling effect,  Phys. Rev. Lett. {\bf 125}, 186802  (2020). 
 \bibitem{LHLi} L. Li, C. H. Lee, S. Mu, and J. Gong, Critical non-Hermitian Skin Effect, Nature communications {\bf 11}, 5491 (2020). 

\bibitem{LeeCH2019} C. H. Lee, L. Li, and J. Gong, Higher-order skin-topological modes in nonreciprocal systems,
Phys. Rev. Lett. {\bf 123}, 016805 (2019).


 \bibitem{Heiss}W. D. Heiss, The physics of exceptional points, J. Phys. A {\bf 45}, 444016 (2012).

 \bibitem{Dembowski}C. Dembowski, B. Dietz, H.-D. Gr\"{a}f, H. L. Harney, A. Heine, W. D. Heiss, and A. Richter, Encircling an exceptional point, Phys. Rev. E {\bf 69}, 056216 (2004).

 \bibitem{Rotter}I. Rotter, A non-Hermitian Hamilton operator and the physics of open quantum systems,  J. Phys. A {\bf 42}, 153001 (2009).

 \bibitem{Kim}J.-W. Ryu, S.-Y. Lee, and S. W. Kim, Analysis of multiple exceptional points related to three interacting eigenmodes in a non-Hermitian Hamiltonian, Phys. Rev. A 85, 042101(2012).
 \bibitem{Hu2017}W. Hu, H. Wang, P. P. Shum, and Y. D. Chong, Exceptional points in a non-Hermitian topological pump, Phys. Rev. B {\bf 95}, 184306 (2017).
 \bibitem{Hassan2017}A. U. Hassan, B. Zhen, M. Soljacic, M. Khajavikhan, and D. N. Christodoulides, Dynamically Encircling Exceptional Points: Exact Evolution and Polarization State Conversion, Phys. Rev. Lett. {\bf 118}, 093002 (2017).

 \bibitem{LeiPan} L. Pan, S. Chen, and X. Cui, High-order exceptional points in ultracold Bose gases, Phys. Rev. A {\bf 99}, 011601(R) (2019).

 \bibitem{Gong} Z. Gong, Y. Ashida, K. Kawabata, K. Takasan, S. Higashikawa,
and M. Ueda, Topological Phases of Non-Hermitian Systems, Phys. Rev. X
\textbf{8}, 031079 (2018).

\bibitem{CHLiu1} C.-H. Liu, H. Jiang, and S. Chen, Topological classification
of non-Hermitian systems with reflection symmetry, Phys. Rev. B \textbf{99},
125103 (2019).

\bibitem{Sato} K. Kawabata, K. Shiozaki, M. Ueda, and M. Sato, Symmetry and
topology in non-Hermitian physics, Phys. Rev. X \textbf{9}, 041015 (2019).

\bibitem{Zhou} H. Zhou and J. Y. Lee, Periodic table for topological bands with non-Hermitian symmetries, Phys. Rev. B \textbf{99}, 235112
(2019).

\bibitem{CHLiu2} C.-H. Liu, and S. Chen, Topological classification of defects
in non-Hermitian systems, Phys. Rev. B \textbf{100}, 144106 (2019).
\bibitem{Ueda}Y. Ashida, Z. Gong, and M. Ueda, Non-Hermitian Physics,  Adv. Phys. \textbf{69}, 249 (2020). 
\bibitem{FSong1}F. Song, S. Yao, and Z. Wang, Non-Hermitian Skin Effect and Chiral Damping in Open Quantum Systems, Phys. Rev. Lett. {\bf 123}, 170401 (2019).

\bibitem{CHLiu3}C.-H. Liu, K. Zhang, Z. Yang, and S. Chen, Helical damping and dynamical critical non-Hermitian skin effect,
                              Phys. Rev. Research, {\bf 2}, 043167 (2020). 




\bibitem{Hodaei}H. Hodaei, A.U. Hassan, S. Wittek, H. Garcia-Gracia, R. El-Ganainy, D.N. Christodoulides, and M. Khajavikhan, Enhanced sensitivity at higher-order exceptional points, Nature (London) {\bf 548}, 187 (2017).
 \bibitem{ChenW} W. Chen, S. K. \"{O}zdemir, G. Zhao, J. Wiersig, and L. Yang, Exceptional points enhance sensing in an optical microcavity, Nature (London) {\bf 548}, 192 (2017).
 \bibitem{Wiersig1}J.  Wiersig, Enhancing the Sensitivity of Frequency and Energy Splitting Detection by Using Exceptional Points: Application to Microcavity Sensors for Single-Particle Detection, Phys. Rev. Lett. {\bf 112}, 203901 (2014).
 \bibitem{Wiersig2}J.  Wiersig, Sensors operating at exceptional points: General theory, Phys. Rev. A {\bf 93}, 033809 (2016).


 \bibitem{IP}The intrinsic property in our paper is defined as the property which does not rely on boundary condition.
 \bibitem{xindex} All $x$ represent cell index in this paper, except that $\sigma_x$ represents the corrseponding Pauli matrix.
\bibitem{nmax} In the classical viewpoint, if $n_x(t)$ takes its maximum value $max(n_x)$ at $t_{max}$, $max(n_x)$ can be regarded as the amplitude of the particle density wave and $t_{max}$ regarded as the time when the wave-front reaches at x, approximatively (regardless the wave length). Thus, $max(n_x)$ can be regarded as the strength of signal and $t_{max}$ regarded as the time when the signal reaches at x (Here, signal represents the particle density wave).
\bibitem{PeskinAndSchroeder} M. E. Peskin and D. V. Schroeder, {\it An Introduction to Quantum Field Theory} (Westview Press, Boulder, 1995).
   \bibitem{helicaltunneling} Assume that $x_a$, $x_b$ are cell indexes, $s_1$, $s_2$, $\bar{s}_1$, $\bar{s}_2$ are spin indexes, $s_1$ and $\bar{s}_1$ represent opposite spin, $s_2$ and $\bar{s}_2$ also represent opposite spin. $T_{x_as_1\rightarrow x_bs_2}$ represents the tunneling amplitude from $x_as_1$ to $x_bs_2$. In analogy to the definition of chiral tunneling \cite{YYi}, the helical tunneling is defined as: $|T_{x_as_1\rightarrow x_bs_2}|\propto \beta_1^{x_a-x_b}$ and $|T_{x_b\bar{s}_1\rightarrow x_a\bar{s}_2}|\propto \beta_1^{x_a-x_b}$, where $\beta_1\ne1$ is a constant.







\bibitem{Haldane}F. D. M. Haldane, Model for a Quantum Hall Effect without Landau Levels: Condensed-Matter Realization of the Parity Anomaly, Phys. Rev. Lett.{\bf 61}, 2015 (1988).
\bibitem{Kane}C. L. Kane and E. J. Mele, Quantum spin hall effect in graphene, Phys. Rev. Lett. {\bf 95}, 226801 (2005).
\bibitem{Bernevig}B. A. Bernevig, T. L. Hughes, and S.-C. Zhang, Quantum Spin Hall Effect and Topological Phase Transition in HgTe Quantum Wells, Science {\bf 314}, 1757 (2006).
\bibitem{Fu}L. Fu, C. L. Kane, and E. J. Mele, Topological Insulators in Three Dimensions, Phys. Rev. Lett. {\bf 98}, 106803(2007).


\bibitem{QLS} The quadratic Lindbladian system in this paper represents the  quadratic fermion Lindbladian system without Copper pairing ($c_m^{\dagger}c_n^{\dagger}$ and $c_mc_n$) terms.




\bibitem{JYLee} J. Y. Lee, J. Ahn, H. Zhou, and A. Vishwanath, Phys. Rev. Lett. {\bf 123}, 206404 (2019).


                                   \end{thebibliography}
\end{document}